\pdfoutput=1

\documentclass[11pt,twoside,a4paper,cmspaper,final,collab]{cms-tdr}
\def\svnVersion{468115P}\def\cmsCernNoTag{CERN-EP-2017-290}\def\cmsCernDate{\today}\def\cmsMessage{Published in the European Physical Journal C as \href{http://dx.doi.org/10.1140/epjc/s10052-018-6033-4}{\doi{10.1140/epjc/s10052-018-6033-4.}}}

\begin{document}\cmsNoteHeader{SMP-16-014}

\hyphenation{had-ron-i-za-tion}
\hyphenation{cal-or-i-me-ter}
\hyphenation{de-vices}
\RCS$Revision: 468112 $
\RCS$HeadURL: svn+ssh://svn.cern.ch/reps/tdr2/papers/SMP-16-014/trunk/SMP-16-014.tex $
\RCS$Id: SMP-16-014.tex 468112 2018-07-10 18:30:47Z panos $

\newlength\cmsFigWidth
\ifthenelse{\boolean{cms@external}}{\setlength\cmsFigWidth{0.98\columnwidth}}{\setlength\cmsFigWidth{0.85\textwidth}}

\ifthenelse{\boolean{cms@external}}{\providecommand{\cmsLeft}{top\xspace}}{\providecommand{\cmsLeft}{left\xspace}}
\ifthenelse{\boolean{cms@external}}{\providecommand{\cmsRight}{bottom\xspace}}{\providecommand{\cmsRight}{right\xspace}}

\providecommand{\alps}{\ensuremath{\alpha_S}\xspace}
\providecommand{\alpsmz}{\ensuremath{\alpha_S(M_{\Z})}\xspace}
\providecommand{\chisq}{\ensuremath{\chi^2}\xspace}
\providecommand{\chisqndof}{\ensuremath{\chi^2/n_\mathrm{dof}}\xspace}
\providecommand{\PYTHIAE} {{\PYTHIA~8}\xspace}
\providecommand{\HERWIGSEVEN} {{\HERWIG~7}\xspace}
\providecommand{\POWHEGTWOJET} {{\textsc{ph-2j}}\xspace}
\providecommand{\POWHEGTWOJETLHE} {{\textsc{ph-2j-lhe}}\xspace}
\providecommand{\POWHEGTHREEJET} {{\textsc{ph-3j}}\xspace}
\providecommand{\RooUnfold} {{\textsc{RooUnfold}}\xspace}
\providecommand{\MGaMCNLO} {{\textsc{MadGraph5\_aMC@NLO}}\xspace}
\providecommand{\NJet} {{\textsc{NJet}}\xspace}
\providecommand{\ptmax}{\ensuremath{\pt^{\text{max}}}\xspace}
\providecommand{\pthat}{\ensuremath{\hat{\text{p}}_{\text{T}}}\xspace}
\providecommand{\mur}{\ensuremath{\mu_r}\xspace}
\providecommand{\muf}{\ensuremath{\mu_f}\xspace}
\providecommand{\rbtrr}{\rule[-0.8ex]{0ex}{3.2ex}}
\providecommand{\met}{\ETslash\xspace}
\providecommand{\metoset}{\ensuremath{\ETslash /\sum{\ET}  }\xspace}
\providecommand{\ptmiss}{\ensuremath{ \vec{p}_\text{T}^\text{miss}}\xspace}

\providecommand{\dphi}{\ensuremath{\Delta\phi_\text{dijet}}\xspace}
\providecommand{\dphiOneTwo}{\ensuremath{\Delta\phi_\text{1,2}}\xspace}
\providecommand{\dphiMinTwoJet}{\ensuremath{\Delta\phi_\text{2j}^\text{min}}\xspace}

\cmsNoteHeader{SMP-16-014}

\title{Azimuthal correlations for inclusive 2-jet, 3-jet, and 4-jet events in pp~collisions at $\sqrt{s}= $ 13\TeV}

\date{\today}

\abstract{	
Azimuthal correlations between
the two jets with the largest transverse momenta \pt in inclusive
2-, 3-, and 4-jet events are presented for several
regions of the leading jet \pt up to 4\TeV.
For 3- and 4-jet scenarios, measurements of
the minimum azimuthal angles between any two
of the three or four leading \pt jets are also presented.
The analysis is based on data from proton-proton collisions
collected by the CMS Collaboration at a centre-of-mass energy of 13\TeV,
corresponding to an integrated luminosity of 35.9\fbinv.
Calculations based on leading-order matrix elements supplemented with
parton showering and hadronization do not fully describe
the data, so next-to-leading-order calculations matched with parton shower and
hadronization models are needed to better describe the measured distributions.
Furthermore, we show that azimuthal jet correlations are sensitive to details of the
parton showering, hadronization, and multiparton interactions.
A next-to-leading-order calculation matched with parton showers in the MC@NLO method, as implemented in \HERWIG~7, gives a better overall description of the measurements than the \POWHEG method.
}

\hypersetup{%
pdfauthor={CMS Collaboration},%
pdftitle={Azimuthal correlations for inclusive 2-jet, 3-jet, and 4-jet events in pp~collisions at sqrt(s)=13 TeV},%
pdfsubject={CMS},%
pdfkeywords={CMS, physics, QCD, jets, azimuthal angle, MC generators}
}

\maketitle

\section{Introduction}
\label{sec:intro}

Particle jets with large transverse momenta \pt are abundantly produced in proton-proton collisions at the CERN LHC
through the strong interactions of quantum chromodynamics (QCD) between the incoming partons.
When the momentum transfer is large, the dynamics can be predicted using perturbative techniques (pQCD).
The two final-state partons at leading order (LO) in pQCD are produced back-to-back in the transverse plane,
and thus the azimuthal angular separation between the two highest-\pt jets,
$\dphiOneTwo=\abs{\phi_\text{jet1}-\phi_\text{jet2}}$, equals~$\pi$.
The production of additional high-\pt jets leads to a deviation of the azimuthal angle from $\pi$.
The measurement of azimuthal angular correlations (or decorrelation from~$\pi$) in inclusive 2-jet
topologies is a useful tool to test theoretical predictions of multijet production processes.
Previous measurements of azimuthal correlation in inclusive 2-jet events were reported by the D0 Collaboration
in $\Pp\PAp$ collisions at $\sqrt{s}=1.96\TeV$ at the Fermilab Tevatron \cite{bib:D0_1,bib:D0_2},
and by the ATLAS Collaboration in $\Pp\Pp$ collisions at $\sqrt{s}=7\TeV$ \cite{bib:ATLAS}
and the CMS Collaboration in $\Pp\Pp$ collisions at $\sqrt{s}=7$ and $8\TeV$ \cite{bib:CMS,bib:CMS_2} at the LHC.
Multijet correlations have been measured by the ATLAS Collaboration at  $\sqrt{s}=8\TeV$ \cite{Aad:2014rma,Aad:2015nda}.

This paper reports measurements of the normalized inclusive 2-, 3-, and 4-jet cross sections
as a function of the azimuthal angular separation between the two highest \pt (leading) jets, \dphiOneTwo,
\begin{equation*}
\frac{1}{\sigma}
\frac{\rd\sigma}{\rd\dphiOneTwo},
\label{eq:jet12}
\end{equation*}
for several regions of the leading jet \pt, \ptmax, for the rapidity region $\abs{y}<2.5$.
The measurements cover the region $\pi/2 < \dphiOneTwo \leq \pi$; the region
$\dphiOneTwo \leq \pi/2$ includes large backgrounds due to \ttbar and Z/W+jet(s) events.
Experimental and theoretical uncertainties are reduced by normalizing the \dphiOneTwo distribution to the
total dijet cross section within each region of \ptmax.

For 3- and 4-jet topologies,  measurements of the normalized inclusive 3- and 4-jet cross sections
are also presented  as a function of the minimum azimuthal angular separation between any two of the
three or four highest \pt jets, \dphiMinTwoJet,
\begin{equation*}
\frac{1}{\sigma}
\frac{\rd\sigma}{\rd\dphiMinTwoJet},
\label{eq:min2j}
\end{equation*}
for several regions of \ptmax, for $\abs{y}<2.5$.
This observable, which is infrared safe (independent of additional soft radiation),
is especially suited for studying correlations amongst the jets in multijet events:
the maximum value of \dphiMinTwoJet is $2 \pi/3$ for 3-jet events (the ``Mercedes star'' configuration),
while it is $\pi/2$ in the 4-jet case (corresponding to the ``cross'' configuration).
The cross section for small angular separations is suppressed because of the finite
jet sizes for a particular jet algorithm.
The observable \dphiMinTwoJet is sensitive to the contributions of jets with lower \pt\ than the leading jet, \ie the subleading jets,
and one can distinguish nearby (nearly collinear) jets (at large \dphiMinTwoJet) from other additional high \pt\ jets (small \dphiMinTwoJet),
yielding information additional to that of the \dphiOneTwo observable.
The 4-jet cross section differential in \dphiMinTwoJet has also been measured by the ATLAS Collaboration \cite{Aad:2015nda}.

The measurements are performed using data collected during 2016 with the CMS experiment at the LHC,
and the event sample corresponds to an integrated luminosity of 35.9\fbinv of proton-proton collisions at $\sqrt{s}=13\TeV$.

\section{The CMS detector}
\label{sec:detector}

The central feature of the CMS detector is a superconducting solenoid, 13\unit{m} in
length and 6\unit{m} in inner diameter, providing an axial magnetic field of 3.8\unit{T}.
Within the solenoid volume are a silicon pixel and strip tracker,
a lead tungstate crystal electromagnetic calorimeter (ECAL)
and a brass and scintillator hadron calorimeter (HCAL), each composed
of a barrel and two endcap sections.
Charged-particle trajectories are measured by the tracker
with full azimuthal coverage within pseudorapidities $\abs{\eta}< 2.5$.
The ECAL, which is equipped with a preshower detector in the endcaps,
and the HCAL cover the region $\abs{\eta}< 3.0$.
Forward calorimeters extend the pseudorapidity coverage provided
by the barrel and endcap detectors to the region $3.0 < \abs{\eta} < 5.2$.
Finally, muons are measured up to $\abs{\eta}< 2.4$ by gas-ionization
detectors embedded in the steel flux-return yoke outside the solenoid.
A detailed description of the CMS detector together with a definition of the coordinate system
used and the relevant kinematic variables can be found in Ref.~\cite{Chatrchyan:2008aa}.

\section{Theoretical predictions}
\label{sec:compMC}

{\tolerance=300
Predictions from five different Monte Carlo (MC) event generators are compared with data.
The \PYTHIAE~\cite{bib:pythia8} and \HERWIGpp~\cite{bib:herwigpp} event generators are used,
both based on LO $2\to2$ matrix element calculations. The \PYTHIAE event generator simulates parton showers ordered
in \pt and uses the Lund string model~\cite{Andersson:1998tv} for hadronization,
while \HERWIGpp generates parton showers through angular-ordered emissions and uses a cluster fragmentation model \cite{Webber:1983if} for hadronization.
The contribution of multiparton interactions (MPI) is simulated in both \PYTHIAE and \HERWIGpp,
but the number of generated MPI varies between \PYTHIAE and \HERWIGpp MPI simulations.
The MPI parameters of both generators are tuned to measurements in proton-proton collisions at the LHC and proton-antiproton
collisions at the Tevatron~\cite{Khachatryan:2015pea}, while the hadronization parameters are determined from fits to LEP data.
For \PYTHIAE the CUETP8M1~\cite{Khachatryan:2015pea} tune, which is based on the NNPDF2.3LO PDF set~\cite{Ball:2013hta,Ball:2011uy}, is employed,
while for  \HERWIGpp the CUETHppS1 tune~\cite{Khachatryan:2015pea}, based on the CTEQ6L1 PDF set~\cite{Pumplin:2002vw}, is used.
\par}

The \MADGRAPH~\cite{bib:madgraph5,Alwall:2007fs} event generator provides LO matrix element calculations with up to
four outgoing partons, \ie $2\to2$, $2\to3$, and $2\to4$ diagrams.
It is interfaced to \PYTHIAE with tune CUETP8M1 for the implementation of parton showers, hadronization, and MPI.
In order to match with \PYTHIAE the \kt-MLM matching procedure \cite{bib:mlm} with a matching scale of $14 \GeV$ is used
to avoid any double counting of the parton configurations generated within the matrix element calculation
and the ones simulated by the parton shower. The NNPDF2.3LO PDF set is used for the hard-process calculation.

{\tolerance=2400
Predictions based on next-to-leading-order (NLO) pQCD are obtained with the \POWHEG {\sc box}
library~\cite{bib:Frixione:2007vw,bib:Alioli:2010xd,bib:Nason:2004rx}
and the \HERWIGSEVEN~\cite{Bellm:2015jjp} event generator.
The events simulated with \POWHEG are matched to \PYTHIAE\ or to \HERWIGpp parton showers and MPI,
while \HERWIGSEVEN uses similar parton shower and MPI models as \HERWIGpp,
and the MC@NLO~\cite{Frixione:2002ik,Frederix:2012ps} method is applied to combine the parton shower with the NLO calculation.
The \POWHEG generator is used in the NLO dijet mode \cite{bib:POWHEG_Dijet}, referred to as \POWHEGTWOJET,
as well as in the NLO three-jet mode~\cite{Kardos:2014dua}, referred to as \POWHEGTHREEJET,
both using the NNPDF3.0NLO PDF set~\cite{Ball:2014uwa}.
The \POWHEG generator, referred to as \POWHEGTWOJETLHE, is also used in the NLO dijet mode without parton showers and MPI.
A minimum \pt for real parton emission of 10 GeV is required for the \POWHEGTWOJET predictions,
and similarly for the \POWHEGTHREEJET predictions a minimum \pt for the three final-state partons of 10 \GeV is imposed.
To simulate the contributions due to parton showers, hadronization, and MPIs, the \POWHEGTWOJET is matched to \PYTHIAE with tune CUETP8M1
and \HERWIGpp with tune CUETHppS1, while the \POWHEGTHREEJET is matched only to \PYTHIAE with tune CUETP8M1.
The matching between the \POWHEG matrix element calculations and the \PYTHIAE underlying event (UE) simulation is performed using
the shower-veto procedure, which rejects showers if their transverse momentum is greater than the minimal \pt of all final-state partons
simulated in the matrix element (parameter \textsc{pthard} = 2~\cite{bib:POWHEG_Dijet}).
Predictions from the \HERWIGSEVEN event generator are based on the MMHT2014 PDF set~\cite{Harland-Lang:2014zoa}
and the default tune H7-UE-MMHT~\cite{Bellm:2015jjp} for the UE simulation.
A summary of the details of the MC event generators used for comparisons with the experimental data is shown in Table~\ref{tableMC}.
\par}

\begin{table*}[htbp]
\begin{center}
\topcaption{Monte Carlo event generators used for comparison in this analysis. Version of the generators, PDF set, underlying event tune, and corresponding references are listed.}
\label{tableMC}
\ifthenelse{\boolean{cms@external}}{}{\resizebox{\textwidth}{!}}
{
\begin{tabular}{c c c c}  \hline
 {Matrix element generator} &  {Simulated diagrams} &  {PDF set} &  {Tune}\\ \hline
 \\[-1.5ex]
 \PYTHIA{}~8.219~\cite{bib:pythia8} &  2$\rightarrow$2 (LO) &  NNPDF2.3LO~\cite{Ball:2013hta,Ball:2011uy} &  CUETP8M1~\cite{Khachatryan:2015pea}
 \\[1.0ex]
\HERWIGpp~2.7.1~\cite{bib:herwigpp} &  2$\rightarrow$2 (LO) &  CTEQ6L1~\cite{Pumplin:2002vw} &  CUETHppS1~\cite{Khachatryan:2015pea}
 \\[1.0ex]
\begin{tabular}[c]{@{}c@{}} \MGaMCNLO 2.3.3~\cite{bib:madgraph5,Alwall:2007fs} \\ +  \PYTHIA{}~8.219~\cite{bib:pythia8} \end{tabular} &  2$\rightarrow$2, 2$\rightarrow$3, 2$\rightarrow$4 (LO) &  NNPDF2.3LO~\cite{Ball:2013hta,Ball:2011uy} &  CUETP8M1~\cite{Khachatryan:2015pea}
\\[2.5ex]
\begin{tabular}[c]{@{}c@{}} \POWHEGTWOJET{}~\footnotesize{V2\_Sep2016}~\cite{bib:Frixione:2007vw,bib:Alioli:2010xd,bib:Nason:2004rx}  \\  +  \PYTHIA{}~8.219~\cite{bib:pythia8}  \end{tabular} &  2$\rightarrow$2 (NLO), 2$\rightarrow$3 (LO) &  NNPDF3.0NLO~\cite{Ball:2014uwa} &  CUETP8M1~\cite{Khachatryan:2015pea}
\\[2.5ex]
\begin{tabular}[c]{@{}c@{}} \POWHEGTWOJETLHE{}~\footnotesize{V2\_Sep2016}~\cite{bib:Frixione:2007vw,bib:Alioli:2010xd,bib:Nason:2004rx}   \end{tabular} &  2$\rightarrow$2 (NLO), 2$\rightarrow$3 (LO) &  NNPDF3.0NLO~\cite{Ball:2014uwa} &
\\[2.5ex]
\begin{tabular}[c]{@{}c@{}} \POWHEGTHREEJET{}~\footnotesize{V2\_Sep2016}~\cite{bib:Frixione:2007vw,bib:Alioli:2010xd,bib:Nason:2004rx}  \\ +  \PYTHIA{}~8.219~\cite{bib:pythia8} \end{tabular} &  2$\rightarrow$3 (NLO), 2$\rightarrow$4 (LO) &  NNPDF3.0NLO~\cite{Ball:2014uwa} &  CUETP8M1~\cite{Khachatryan:2015pea}
\\[2.5ex]
\begin{tabular}[c]{@{}c@{}} \POWHEGTWOJET{}~\footnotesize{V2\_Sep2016}~\cite{bib:Frixione:2007vw,bib:Alioli:2010xd,bib:Nason:2004rx} \\  + \HERWIGpp~2.7.1~\cite{bib:herwigpp}  \end{tabular}  &  2$\rightarrow$2 (NLO), 2$\rightarrow$3 (LO) &  NNPDF3.0NLO~\cite{Ball:2014uwa} &  CUETHppS1~\cite{Khachatryan:2015pea}
\\[2.5ex]
\HERWIG~7.0.4~\cite{Bellm:2015jjp} &  2$\rightarrow$2 (NLO), 2$\rightarrow$3 (LO) &  MMHT2014~\cite{Harland-Lang:2014zoa}  &  H7-UE-MMHT~\cite{Bellm:2015jjp}
\\[1.0ex]
\hline
\end{tabular}
}
\end{center}
\end{table*}

Uncertainties in the theoretical predictions of the parton shower simulation are illustrated using
the \PYTHIAE event generator. Choices of scale for the parton shower are expected to
have the largest impact on the azimuthal distributions.
The parton shower uncertainty is calculated by independently varying
the renormalization scales ($\mu_r$) for initial- and final-state radiation by a factor 2 in units
of the \pt of the emitted partons of the hard scattering.
The maximum deviation found is considered a theoretical uncertainty in the event generator predictions.

\section{Jet reconstruction and event selection}
\label{sec:selectionReconstruction}

The measurements are based on data samples collected with single-jet high-level triggers
(HLT) \cite{CMS_HLT,CMS_TRG}. Five such triggers are considered that require
at least one jet in an event with $\pt > 140$, 200, 320, 400, or $450\GeV$
in the full rapidity coverage of the CMS detector.
All triggers are prescaled except the one with the highest threshold.
Table~\ref{tbl:TrigLumi} shows the integrated luminosity \lumi for the five trigger samples.
The relative efficiency of each trigger is estimated using triggers with lower \pt thresholds.
Using these five jet energy thresholds, a 100\% trigger efficiency is achieved
in the region of $\ptmax > 200\GeV $.

\begin{table*}[htb]
\centering
\topcaption{The integrated luminosity for each trigger
sample considered in this analysis.\label{tbl:TrigLumi}}
\begin{tabular}{lccccc}
\hline
HLT \pt threshold (\GeVns{})   &  140    &  200    &  320  & 400  &  450 \\
\ptmax region (\GeVns{})   &  200--300   & 300--400    &  400--500  & 500--600  &  $>$600 \\
\hline
\lumi ($\text{fb}^{-1}$)\rule[-0.8ex]{0ex}{3.2ex} &  0.024  &  0.11  &  1.77 &  5.2 & 36 \\
\hline
\end{tabular}
\end{table*}

{\tolerance=1200
Particles are reconstructed and identified using a particle-flow (PF) algorithm~\cite{CMS-PRF-14-001},
which uses an optimized combination of information from the various elements of the CMS detector.
Jets are reconstructed by clustering the Lorentz vectors of the PF candidates with the infrared-
and collinear-safe anti-\kt clustering algorithm \cite{bib:antikt} with a distance parameter $R=0.4$.
The clustering is performed with the \FASTJET package \cite{bib:fastjet}.
The technique of charged-hadron subtraction \cite{bib:jes8} is used to remove tracks identified as originating
from additional pp interactions within the same or neighbouring bunch crossings (pileup).
The average number of pileup interactions observed in the data is about 27.
\par}

The reconstructed jets require energy corrections to account for residual nonuniformities and
nonlinearities in the detector response.
These jet energy scale (JES) corrections \cite{bib:jes8} are derived using simulated events that are generated
with \PYTHIA~8.219 \cite{bib:pythia8} using tune CUETP8M1 \cite{Khachatryan:2015pea} and processed through
the CMS detector simulation based on \GEANTfour \cite{bib:geant}; they are confirmed with in situ measurements with dijet,
multijet, photon+jet, and leptonic \textit{Z}+jet events.
An offset correction is required to account for the extra energy clustered into jets due to pileup.
The JES corrections, which depend on the $\eta$ and \pt of the jet, are applied as multiplicative
factors to the jet four-momentum vectors. The typical overall correction is about 10\% for central jets
having $\pt = 100\GeV$ and decreases with increasing \pt.

{\tolerance=1200
Resolution studies on the measurements of \dphiOneTwo and \dphiMinTwoJet are performed using \PYTHIA~8.219
with tune CUETP8M1 processed through the CMS detector simulation.
The azimuthal angular separation is determined with an accuracy from $1^\circ$ to $0.5^\circ$ (0.017 to 0.0087 in radians)
for $\ptmax = 200\GeV$ to $1\TeV$, respectively.
\par}

Events are required to have at least one primary vertex candidate \cite{bib:vertex_paper} reconstructed
offline from at least five charged-particle tracks and lies along the beam line within 24\unit{cm} of the
nominal interaction point.
The reconstructed vertex with the largest value of summed physics-object $\pt^2$ is taken to be
the primary $\Pp\Pp$ interaction vertex. The physics objects are the objects determined by a jet
finding algorithm~\cite{bib:antikt,bib:fastjet} applied to all charged tracks associated with the vertex
plus the corresponding associated missing transverse momentum.
Additional selection criteria are applied to each event to remove spurious jet-like signatures originating
from isolated noise patterns in certain HCAL regions.
Stringent criteria \cite{bib:jetID_PF} are applied to suppress these nonphysical signatures;
each jet should contain at least two particles, one of which is a charged hadron,
and the jet energy fraction carried by neutral hadrons and photons should be less than 90\%.
These criteria have a jet selection efficiency greater than 99\% for genuine jets.

For the measurements of the normalized inclusive \mbox{2-}, \mbox{3-}, and \mbox{4-jet} cross sections
as a function of \dphiOneTwo or \dphiMinTwoJet all jets in the event with $\pt > 100 \GeV$ and a
rapidity $\abs{y}<5$ are considered and ordered in \pt.
Events are selected where the two highest-\pt jets have $\abs{y}<2.5$,
(\ie events are not counted where one of the leading jets has $\abs{y}>2.5$).
Also, events are only selected in which the highest-\pt jet has $\abs{y}<2.5$ and exceeds 200\GeV.
The inclusive 2-jet event sample includes events where the two
leading jets lie within the tracker coverage of $\abs{y}<2.5$.
Similarly the 3-jet (4-jet) event sample includes those events where
the three (four) leading jets lie within $\abs{y}<2.5$, respectively.
In this paper results are presented in bins of \ptmax, corresponding to the \pt
of the leading jet, which is always within $\abs{y}<2.5$.

\section{\texorpdfstring{Measurements of the normalized inclusive 2-, 3-, and 4-jet cross sections in \dphiOneTwo and \dphiMinTwoJet}
{Measurements of the normalized inclusive 2-, 3-, and 4-jet cross sections in dphiOneTwo and dphiMinTwoJet}}
\label{sec:measurement}

The normalized inclusive 2-, 3-, and 4-jet cross sections differential in \dphiOneTwo and \dphiMinTwoJet
are corrected for the finite detector resolution to better approximate the final-state particles,
a procedure called "unfolding".
In this way, a direct comparison of this measurement to results from other experiments and to QCD predictions is possible.
Particles are considered stable if their mean decay length is $c\tau > 1\unit{cm}$.

The bin width used in the measurements of \dphiOneTwo and \dphiMinTwoJet is set to $\pi / 36 = 0.087$ rads ($5^\circ$),
which is five to ten times larger than the azimuthal angular separation resolution.
The corrections due to the unfolding are approximately a few per cent.

The unfolding procedure is based on the matrix inversion algorithm implemented in the software package
\RooUnfold \cite{bib:RooUnfold} using a 2-dimensional response matrix that correlates the modeled distribution with the reconstructed one.
The response matrix is created by the convolution of the $\Delta\phi$ resolution with the generator-level
inclusive 2-, 3-, and 4- cross section distributions from \PYTHIAE with tune CUETP8M1.
The unfolded distributions differ from the distributions at detector level by 1 to 4\%.
As a cross-check, the above procedure was repeated by creating the response matrix with event samples
obtained with the full \GEANTfour detector simulation, and no significant difference was observed.

We consider three main sources of systematic uncertainties that arise from the estimation of the JES
calibration, the jet energy resolution (JER), and the unfolding correction.
The relative JES uncertainty is estimated to be 1--2\% for PF jets using charged-hadron subtraction \cite{bib:jes8}.
The resulting uncertainties in the normalized 2-, 3-, and 4-jet cross sections differential in \dphiOneTwo
range from 3\% at $\pi/2$ to 0.1\% at $\pi$.
For the normalized 3- and 4-jet cross sections differential in \dphiMinTwoJet the resulting uncertainties
range from 0.1 to 1\%, and 0.1 to 2\%, respectively.

The JER~\cite{bib:jes8} is responsible for migration of events among the \ptmax regions, and its parametrization
is determined from a full detector simulation using events generated by \PYTHIAE with tune CUETP8M1.
The effect of the JER uncertainty is estimated by varying its parameters within their uncertainties \cite{bib:jes8}
and comparing the normalized inclusive 2-, 3-, and 4-jet cross sections before and after the changes.
The JER-induced uncertainty ranges from 1\% at $\pi/2$ to 0.1\% at $\pi$ for the normalized 2-, 3-, and 4-jet cross
sections differential in \dphiOneTwo and is less than 0.5\% for the normalized 3- and 4-jet cross sections
differential in \dphiMinTwoJet.

The above systematic uncertainties in the JES calibration and the JER cover the effects from migrations due to the
\pt\ thresholds, \ie migrations between the \mbox{2-}, \mbox{3-}, and \mbox{4-jet} samples
and migrations between the various \ptmax regions of the measurements.

The unfolding procedure is affected by uncertainties in the parametrization of the $\Delta\phi$ resolution.
Alternative response matrices, generated by varying the $\Delta\phi$ resolution by $\pm$10\%,
are used to unfold the measured spectra.
This variation is motivated by studies on the $\Delta\phi$ resolution for simulated di-jet events~\cite{CMS-PRF-14-001}.
The uncertainty in the unfolding correction factors is estimated to be about 0.2\%.
An additional systematic uncertainty is obtained by examining the dependence of the response matrix
on the choice of the MC generator. Alternative response matrices are constructed using the
\HERWIGpp event generator \cite{bib:herwigpp} with tune EE5C~\cite{Seymour:2013qka}; the effect is $<$0.1\%.
A total systematic unfolding uncertainty of 0.2\% is considered, which accounts for all these various uncertainty sources.

\section{Comparison with theoretical predictions}

\subsection{\texorpdfstring{The \dphiOneTwo measurements}{The dphiOneTwo measurements}}

\begin{figure}[hbtp]
\centering
\includegraphics[width=\cmsFigWidth]{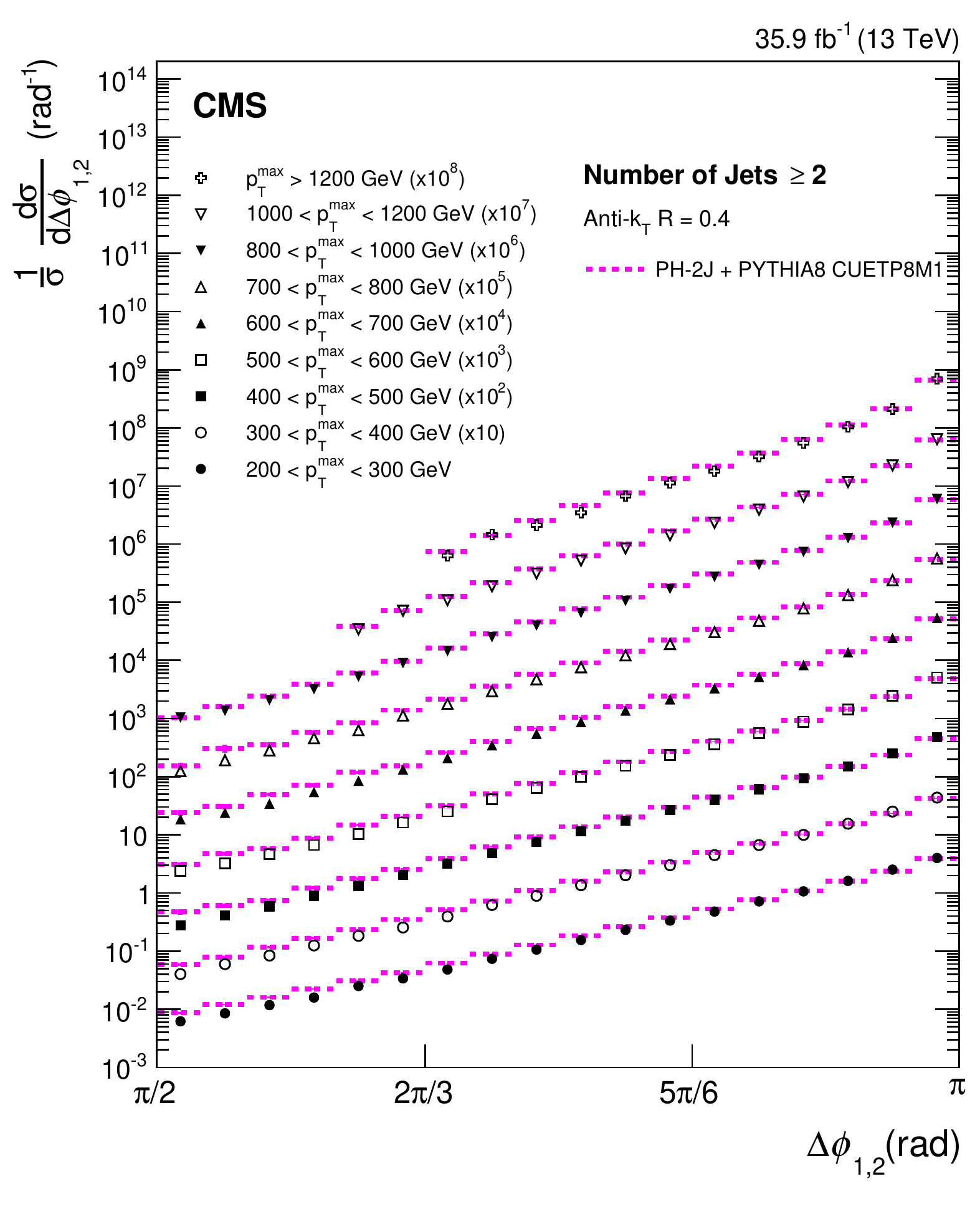}
\caption{Normalized inclusive 2-jet cross section differential in \dphiOneTwo
for nine \ptmax regions, scaled by multiplicative factors for
presentation purposes.
The size of the data symbol includes both statistical and systematic uncertainties.
The data points are overlaid with the predictions from
the \POWHEGTWOJET + \PYTHIAE  event generator.}
\label{fig:2J_particle_xsection_MC_data_12}
\end{figure}

\begin{figure}[hbtp]
\centering
\includegraphics[width=\cmsFigWidth]{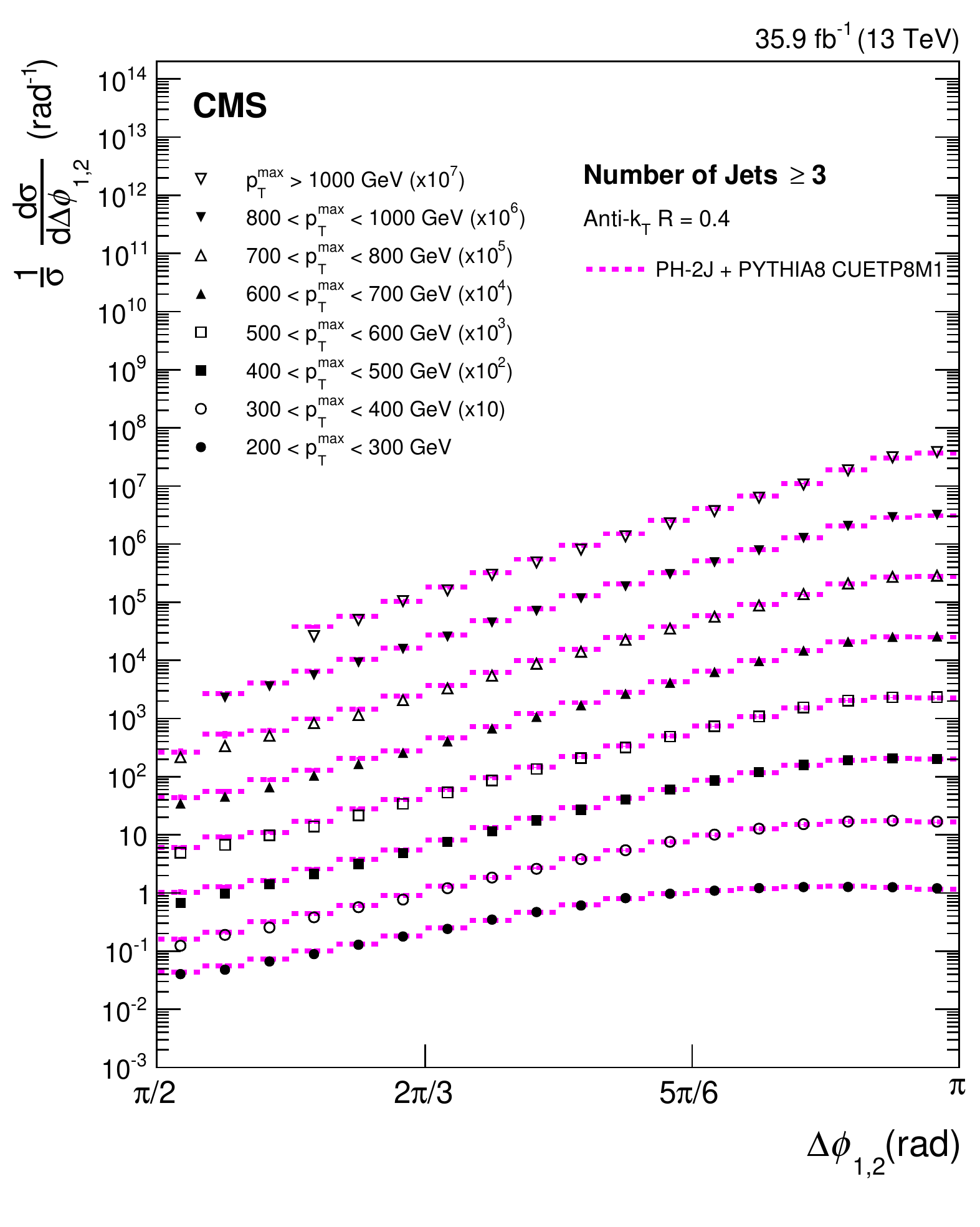}
\caption{Normalized inclusive 3-jet cross section differential in \dphiOneTwo
for eight \ptmax regions, scaled by multiplicative factors for
presentation purposes.
The size of the data symbol includes both statistical and systematic uncertainties.
The data points are overlaid with the predictions from
the \POWHEGTWOJET + \PYTHIAE  event generator.}
\label{fig:3J_particle_xsection_MC_data_12}
\end{figure}

\begin{figure}[hbtp]
\centering
\includegraphics[width=\cmsFigWidth]{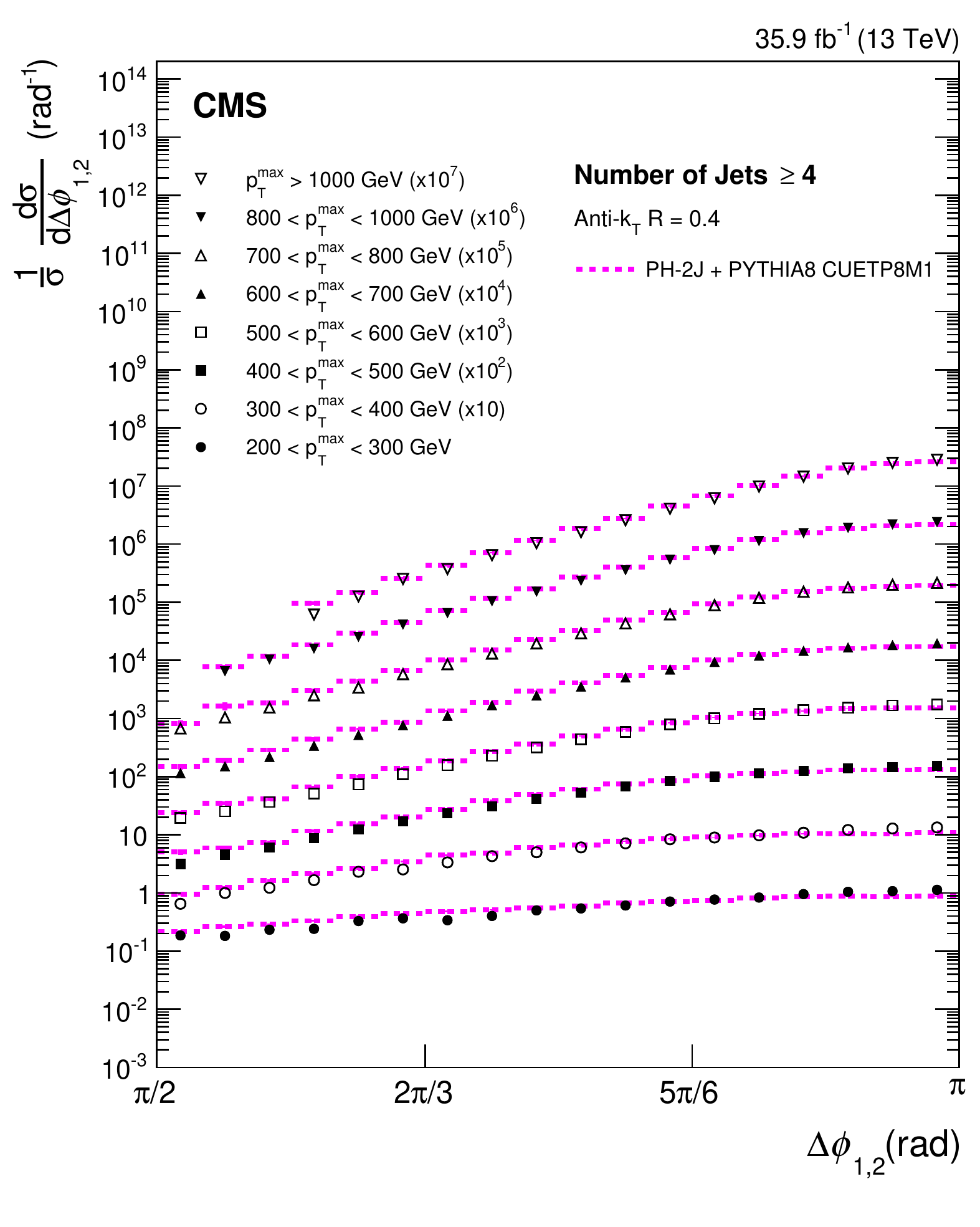}
\caption{Normalized inclusive 4-jet cross section differential in \dphiOneTwo
for eight \ptmax regions, scaled by multiplicative factors for
presentation purposes.
The size of the data symbol includes both statistical and systematic uncertainties.
The data points are overlaid with the predictions from
the \POWHEGTWOJET + \PYTHIAE  event generator.}
\label{fig:4J_particle_xsection_MC_data_12}
\end{figure}

The unfolded, normalized, inclusive 2-, 3-, and 4-jet cross sections differential in \dphiOneTwo are shown in
Figs.~\ref{fig:2J_particle_xsection_MC_data_12}-\ref{fig:4J_particle_xsection_MC_data_12}
for the various \ptmax regions considered in this analysis.
In the 2-jet case the \dphiOneTwo distributions are strongly peaked at $\pi$ and become steeper with increasing \ptmax.
In the 3-jet case, the \dphiOneTwo distributions become flatter at $\pi$, since by definition dijet events do not contribute,
and in the 4-jet case they become even flatter.
The data points are overlaid with the predictions from the \POWHEGTWOJET + \PYTHIAE  event generator.

\begin{figure}[hbtp]
\centering
\includegraphics[width=\cmsFigWidth]{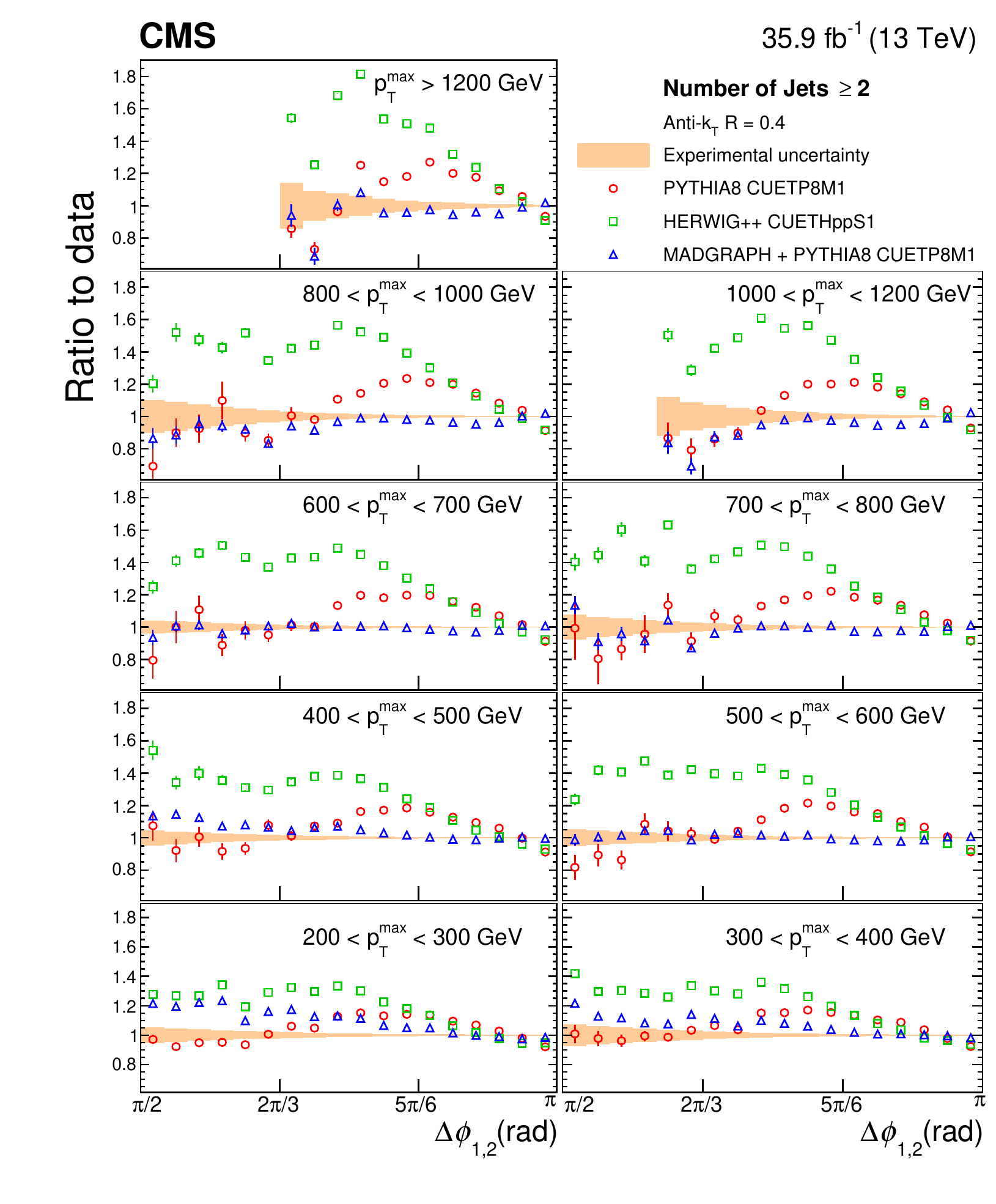}
\caption{Ratios of \PYTHIAE, \HERWIGpp, and \MADGRAPH + \PYTHIAE predictions to the normalized
inclusive 2-jet cross section differential in \dphiOneTwo, for all \ptmax regions.
The solid band indicates the total experimental uncertainty and
the vertical bars on the points represent the statistical
uncertainties in the simulated data.}
\label{fig:2J_ratios_MC_data_a_12}
\end{figure}

\begin{figure}[hbtp]
\centering
\includegraphics[width=\cmsFigWidth]{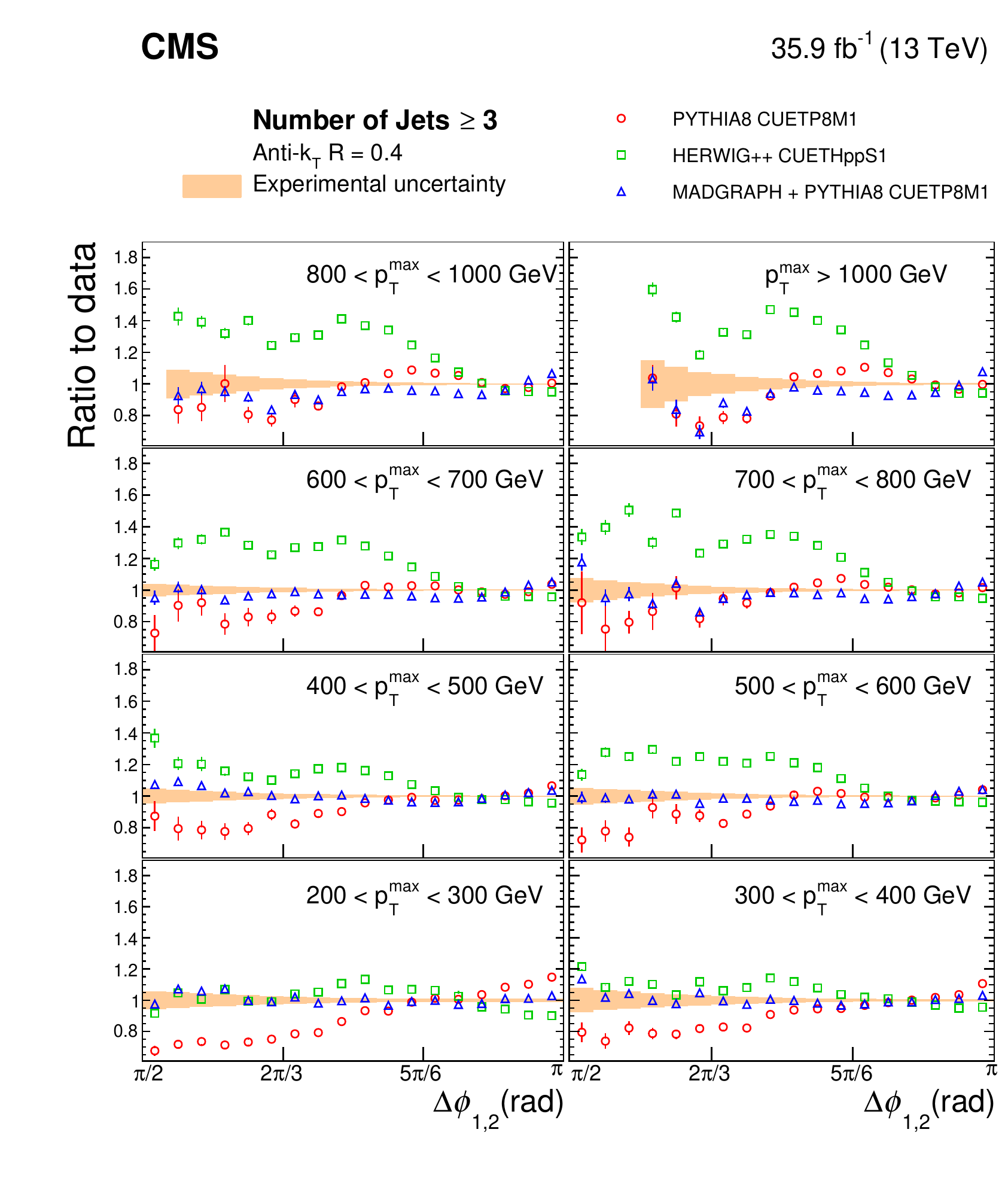}
\caption{Ratios of \PYTHIAE, \HERWIGpp, and \MADGRAPH + \PYTHIAE predictions to the normalized
inclusive 3-jet cross section differential in \dphiOneTwo, for all \ptmax regions.
The solid band indicates the total experimental uncertainty and
the vertical bars on the points represent the statistical
uncertainties in the simulated data.}
\label{fig:3J_ratios_MC_data_a_12}
\end{figure}

\begin{figure}[hbtp]
\centering
\includegraphics[width=\cmsFigWidth]{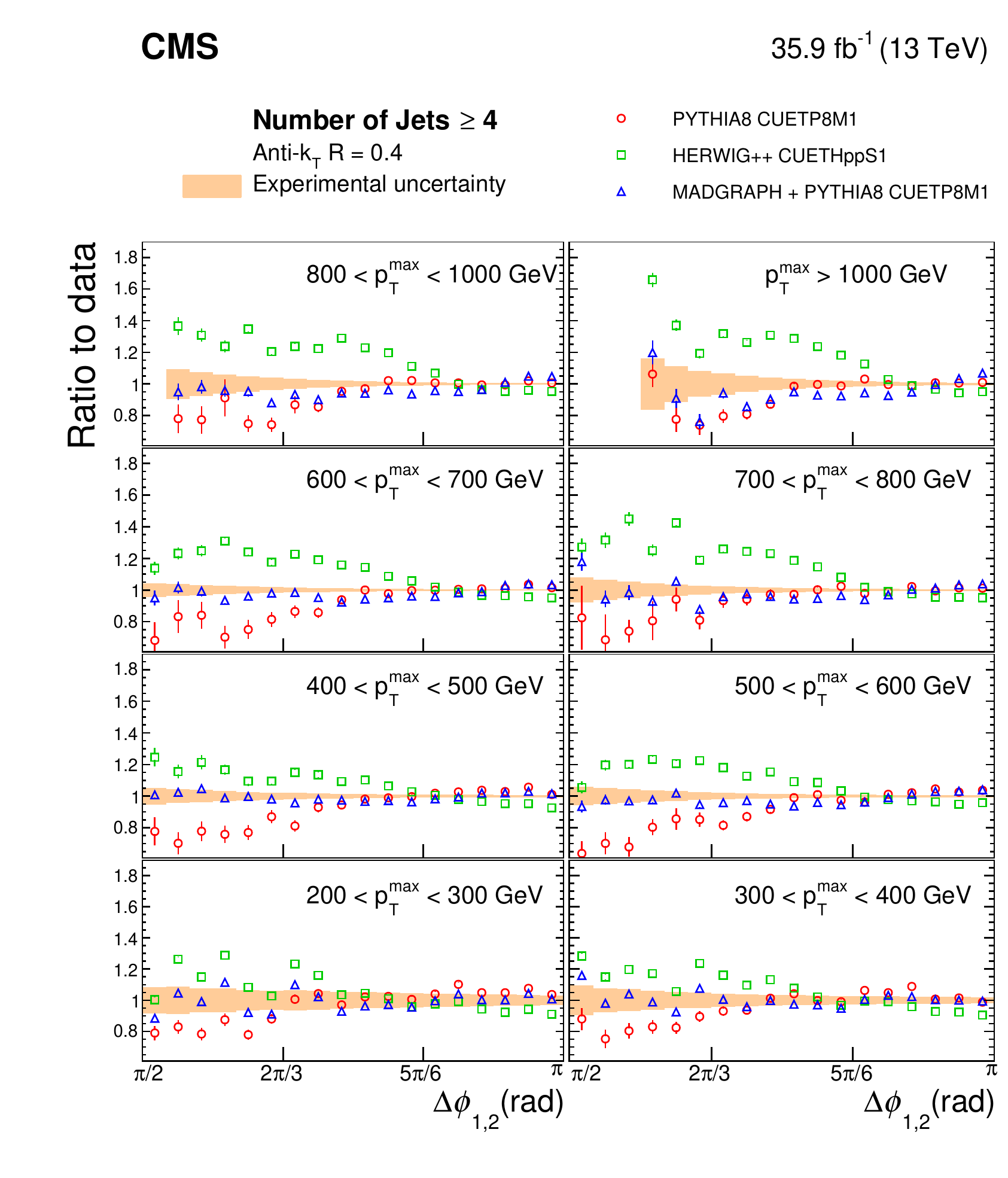}
\caption{Ratios of \PYTHIAE, \HERWIGpp, and \MADGRAPH + \PYTHIAE predictions to the normalized
inclusive 4-jet cross section differential in \dphiOneTwo, for all \ptmax regions.
The solid band indicates the total experimental uncertainty and
the vertical bars on the points represent the statistical
uncertainties in the simulated data.}
\label{fig:4J_ratios_MC_data_a_12}
\end{figure}

The ratios of the \PYTHIAE, \HERWIGpp, and \MADGRAPH + \PYTHIAE event generator predictions to the normalized inclusive 2-,
3-, and 4-jet cross section differential in \dphiOneTwo are shown in
Figs.~\ref{fig:2J_ratios_MC_data_a_12}, \ref{fig:3J_ratios_MC_data_a_12}, and~\ref{fig:4J_ratios_MC_data_a_12},
respectively, for all \ptmax regions.
The solid band around unity represents the total experimental uncertainty and the error
bars on the points represent the statistical uncertainties in the simulated data.
Among the LO dijet event generators, \HERWIGpp exhibits the largest deviations from the experimental measurements,
whereas \PYTHIAE behaves much better than \HERWIGpp,
although with deviations of up to 30-40\%, in particular around $ \dphiOneTwo = 5\pi/6$ in the 2-jet case
and around $\dphiOneTwo < 2\pi/3$ in the 3- and 4-jet case.
Predictions from \HERWIGpp tend to overestimate the measurements as a function of  $ \dphiOneTwo$  in the
2-, 3-, and 4-jet cases, especially at  $ \dphiOneTwo < 5\pi/6$  for $\ptmax > 400 \GeV$.
However, it is remarkable that predictions based on the $2\to 2$ matrix element calculations supplemented with parton showers, MPI, and hadronization describe
the $\dphiOneTwo$ distributions rather well, even in regions that are sensitive to hard jets not included in the matrix element calculations.
The \MADGRAPH + \PYTHIAE calculation using up to 4 partons in the matrix element calculations provides the best description of the measurements.

\begin{figure}[hbtp]
\centering
\includegraphics[width=\cmsFigWidth]{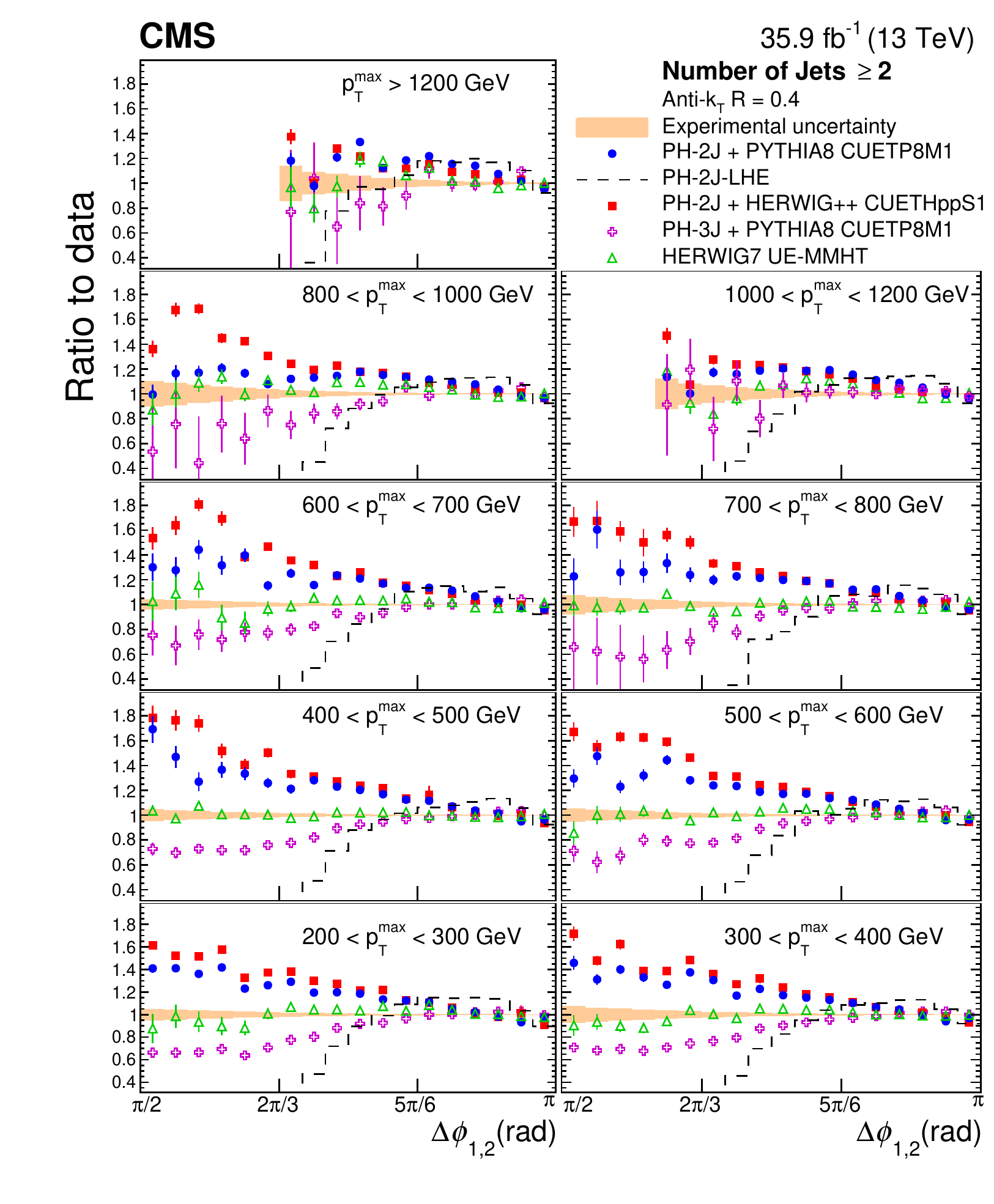}
\caption{Ratios of  \POWHEGTWOJET + \PYTHIAE, \POWHEGTWOJETLHE, \POWHEGTWOJET + \HERWIGpp,
\POWHEGTHREEJET + \PYTHIAE, and \HERWIGSEVEN predictions to the normalized
inclusive 2-jet cross section differential in \dphiOneTwo, for all \ptmax regions.
The solid band indicates the total experimental uncertainty and
the vertical bars on the points represent the statistical
uncertainties in the simulated data.}
\label{fig:2J_ratios_MC_data_b_12}
\end{figure}

\begin{figure}[hbtp]
\centering
\includegraphics[width=\cmsFigWidth]{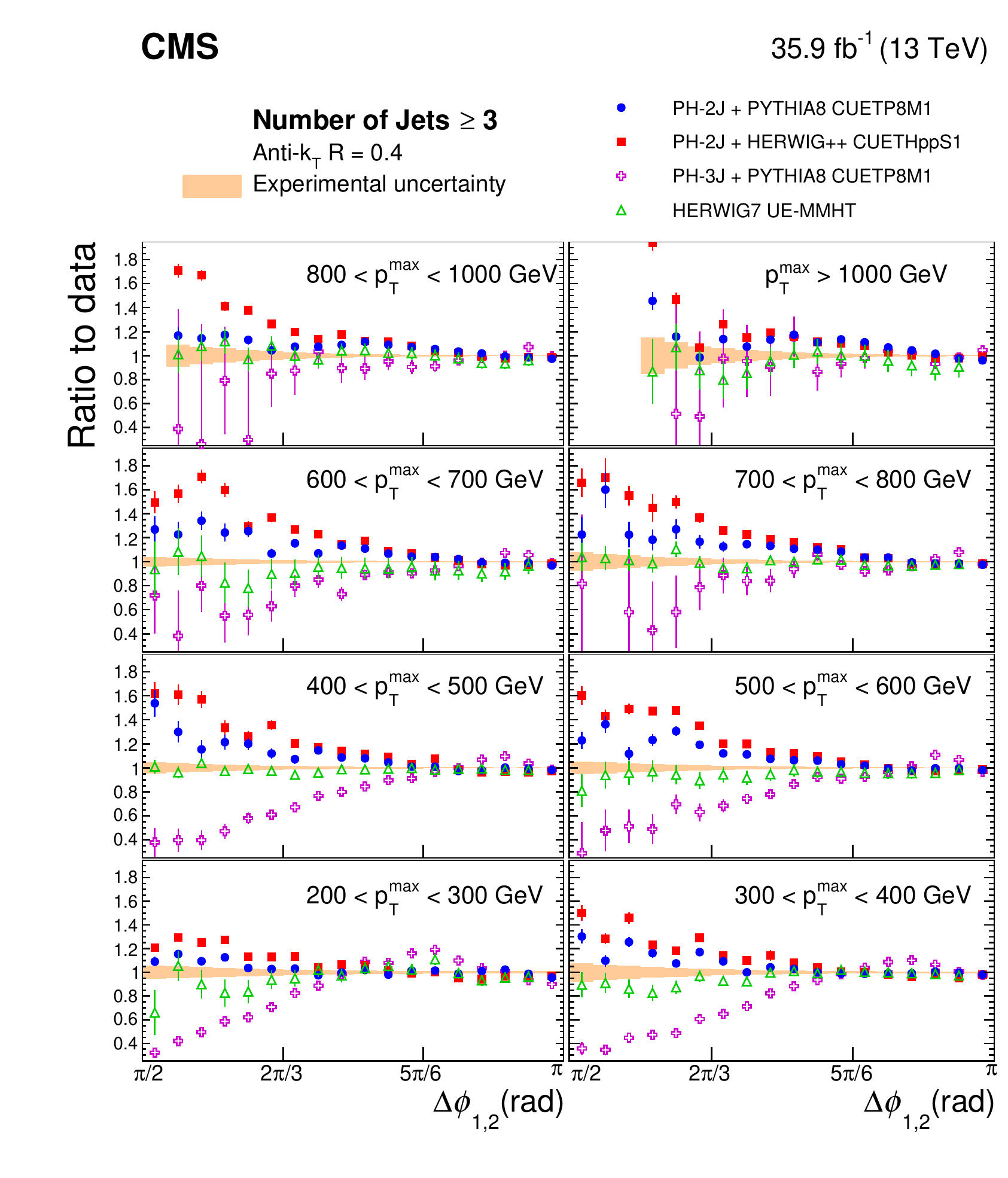}
\caption{Ratios of  \POWHEGTWOJET + \PYTHIAE, \POWHEGTWOJET + \HERWIGpp,
\POWHEGTHREEJET + \PYTHIAE, and \HERWIGSEVEN predictions to the normalized
inclusive 3-jet cross section differential in \dphiOneTwo, for all \ptmax regions.
The solid band indicates the total experimental uncertainty and
the vertical bars on the points represent the statistical
uncertainties in the simulated data.}
\label{fig:3J_ratios_MC_data_b_12}
\end{figure}

\begin{figure}[hbtp]
\centering
\includegraphics[width=\cmsFigWidth]{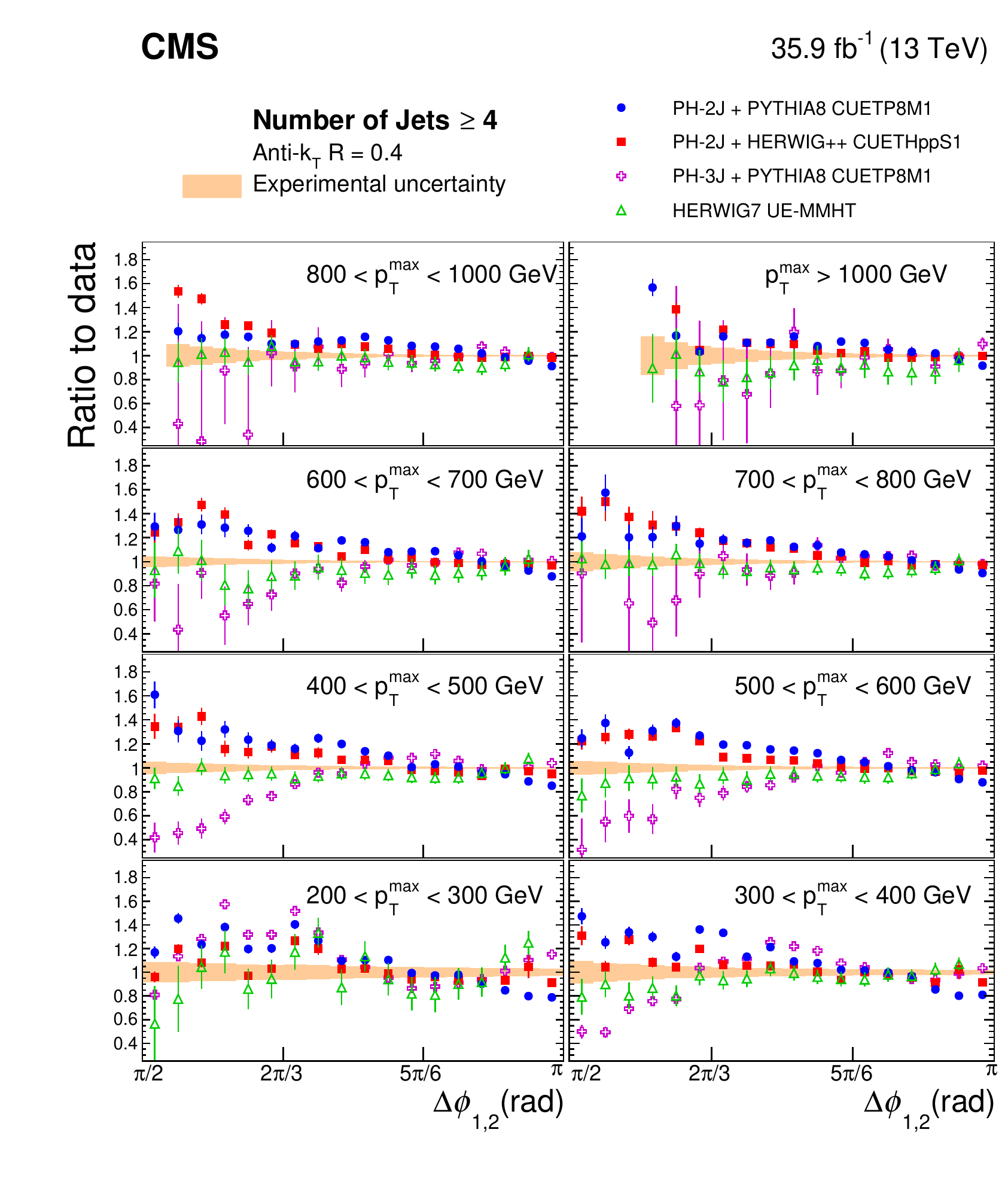}
\caption{Ratios of  \POWHEGTWOJET + \PYTHIAE, \POWHEGTWOJET + \HERWIGpp,
\POWHEGTHREEJET + \PYTHIAE, and \HERWIGSEVEN predictions to the normalized
inclusive 4-jet cross section differential in \dphiOneTwo, for all \ptmax regions.
The solid band indicates the total experimental uncertainty and
the vertical bars on the points represent the statistical
uncertainties in the simulated data.}
\label{fig:4J_ratios_MC_data_b_12}
\end{figure}

Figures~\ref{fig:2J_ratios_MC_data_b_12}-\ref{fig:4J_ratios_MC_data_b_12}
show the ratios of the \POWHEGTWOJET matched to \PYTHIAE and \HERWIGpp, \POWHEGTHREEJET + \PYTHIAE,
and \HERWIGSEVEN event generators predictions to the normalized inclusive 2-,
3-, and 4-jet cross section differential in \dphiOneTwo, for all \ptmax regions.
The solid band around unity represents the total experimental uncertainty and the vertical
bars on the points represent the statistical uncertainties in the simulated data.
The predictions of \POWHEGTWOJET and \POWHEGTHREEJET exhibit deviations from the measurement, increasing towards small \dphiOneTwo.
While \POWHEGTWOJET is above the data, \POWHEGTHREEJET predicts too few events at small \dphiOneTwo.
These deviations were investigated in a dedicated study with parton showers and MPI switched off.
Because of the kinematic restriction of a 3-parton state, \POWHEGTWOJET without parton showers cannot fill the region
$\dphiOneTwo <2\pi/3$, shown as \POWHEGTWOJETLHE with the dashed line in Fig.~\ref{fig:2J_ratios_MC_data_b_12},
whereas for \POWHEGTHREEJET the parton showers have little impact.
Thus, the events at low \dphiOneTwo observed for \POWHEGTWOJET originate from leading-log parton showers, and there are too many of these.
In contrast, the \POWHEGTHREEJET prediction, which provides $2\to 3$ jet calculations at NLO QCD, is below the measurement.
The NLO \POWHEGTWOJET calculation and the LO \POWHEG three-jet calculation are
equivalent when initial- and final-state radiation are not allowed to occur.

The predictions from \POWHEGTWOJET matched to \PYTHIAE describe the normalized cross sections better than those where \POWHEGTWOJET
is matched to \HERWIGpp.
Since the hard process calculation is the same, the difference between the two predictions might be due to the treatment
of parton showers in \PYTHIAE and \HERWIGpp and to the matching to the matrix element calculation.
The \PYTHIAE and \HERWIGpp parton shower calculations use different $\alpha_S$ values for initial- and final-state emissions,
in addition to a different upper scale for the parton shower simulation, which is higher in \PYTHIAE than in \HERWIGpp.
The dijet NLO calculation of \HERWIGSEVEN provides the best description of the measurements,
indicating that the MC@NLO method of combining parton showers with the NLO parton level calculations has advantages compared to the POWHEG method
in this context.

For $\dphiOneTwo$ generator-level predictions in the 2-jet case, parton shower uncertainties have a very small
impact ($<$5\%) at values close to $\pi$ and go up to 40--60\% for increasing $\ptmax$ at $\dphiOneTwo\sim\pi/2$.
For the 3- and 4-jet scenarios, parton shower uncertainties are less relevant, not exceeding $\sim$20\% for $\dphiOneTwo$.

\subsection{\texorpdfstring{The \dphiMinTwoJet measurements}{The dphiMinTwoJet measurements}}

\begin{figure}[hbtp]
\centering
\includegraphics[width=\cmsFigWidth]{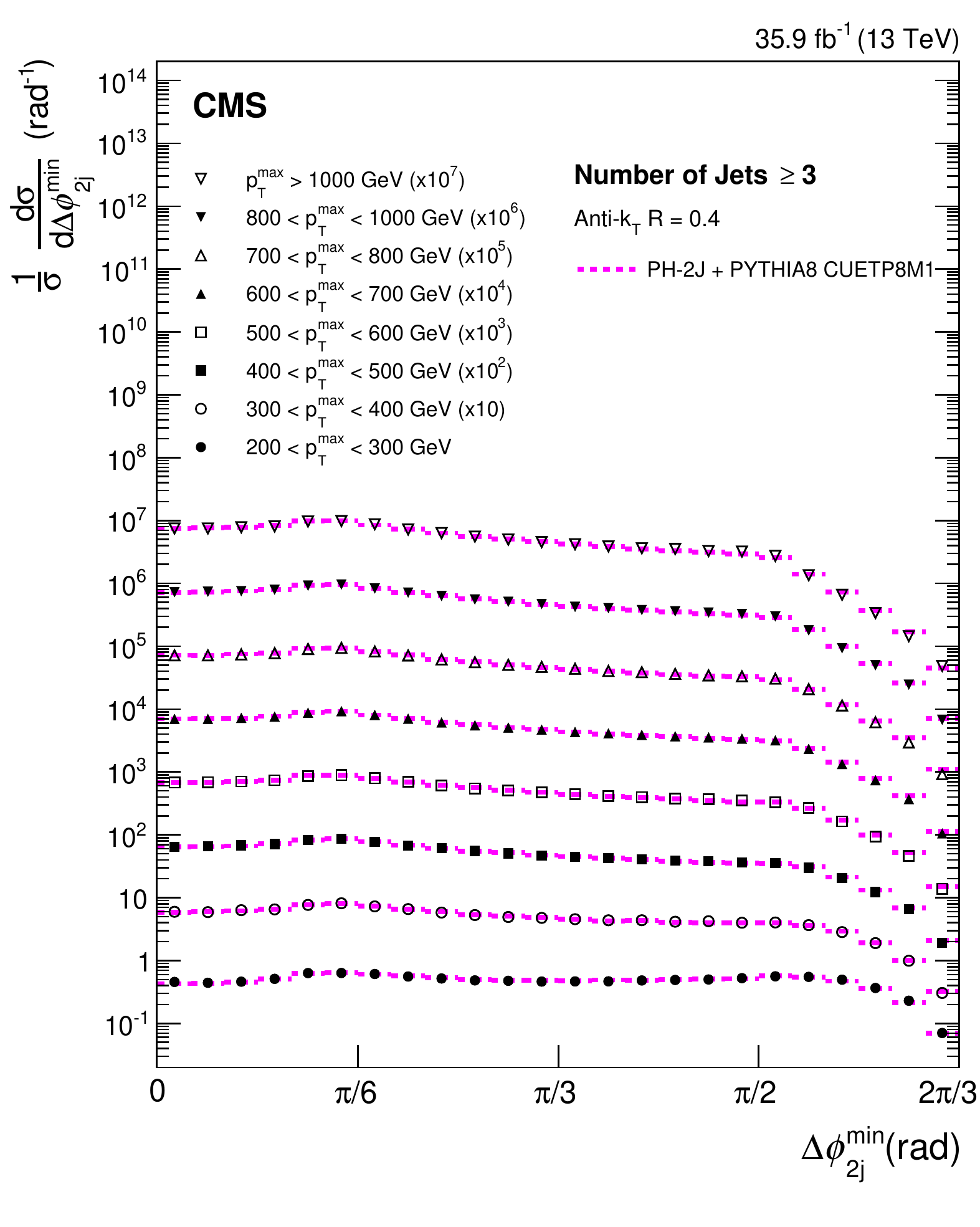}
\caption{Normalized inclusive 3-jet cross section differential in \dphiMinTwoJet
for eight \ptmax regions, scaled by multiplicative factors for
presentation purposes.
The size of the data symbol includes both statistical and systematic uncertainties.
The data points are overlaid with the predictions from
the \POWHEGTWOJET + \PYTHIAE  event generator.}
\label{fig:3J_particle_xsection_MC_data_min2j}
\end{figure}

\begin{figure}[hbtp]
\centering
\includegraphics[width=\cmsFigWidth]{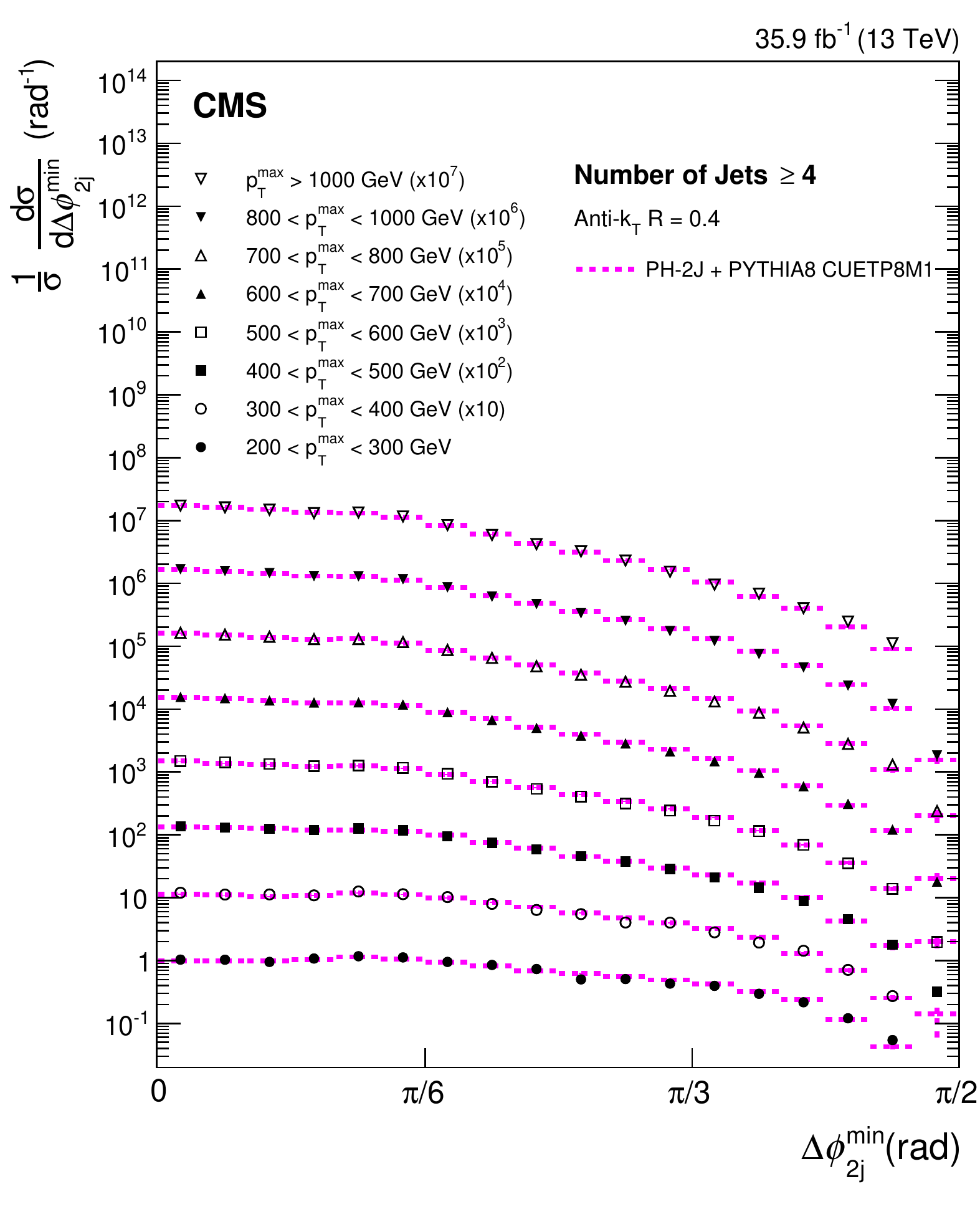}
\caption{Normalized inclusive 4-jet cross section differential in \dphiMinTwoJet
for eight \ptmax regions, scaled by multiplicative factors for
presentation purposes.
The size of the data symbol includes both statistical and systematic uncertainties.
The data points are overlaid with the predictions from
the \POWHEGTWOJET + \PYTHIAE  event generator.}
\label{fig:4J_particle_xsection_MC_data_min2j}
\end{figure}

The unfolded, normalized, inclusive 3- and 4-jet cross sections differential in \dphiMinTwoJet are shown in
Figs.~\ref{fig:3J_particle_xsection_MC_data_min2j} and \ref{fig:4J_particle_xsection_MC_data_min2j}, respectively, for eight \ptmax regions.
The measured distributions decrease towards the kinematic limit of $\dphiMinTwoJet
\to 2 \pi/3( \pi/2)$ for the 3-jet and 4-jet case, respectively.
The data points are overlaid with the predictions from the \POWHEGTWOJET + \PYTHIAE  event generator.
The size of the data symbol includes both statistical and systematic uncertainties.

\begin{figure}[hbtp]
\centering
\includegraphics[width=\cmsFigWidth]{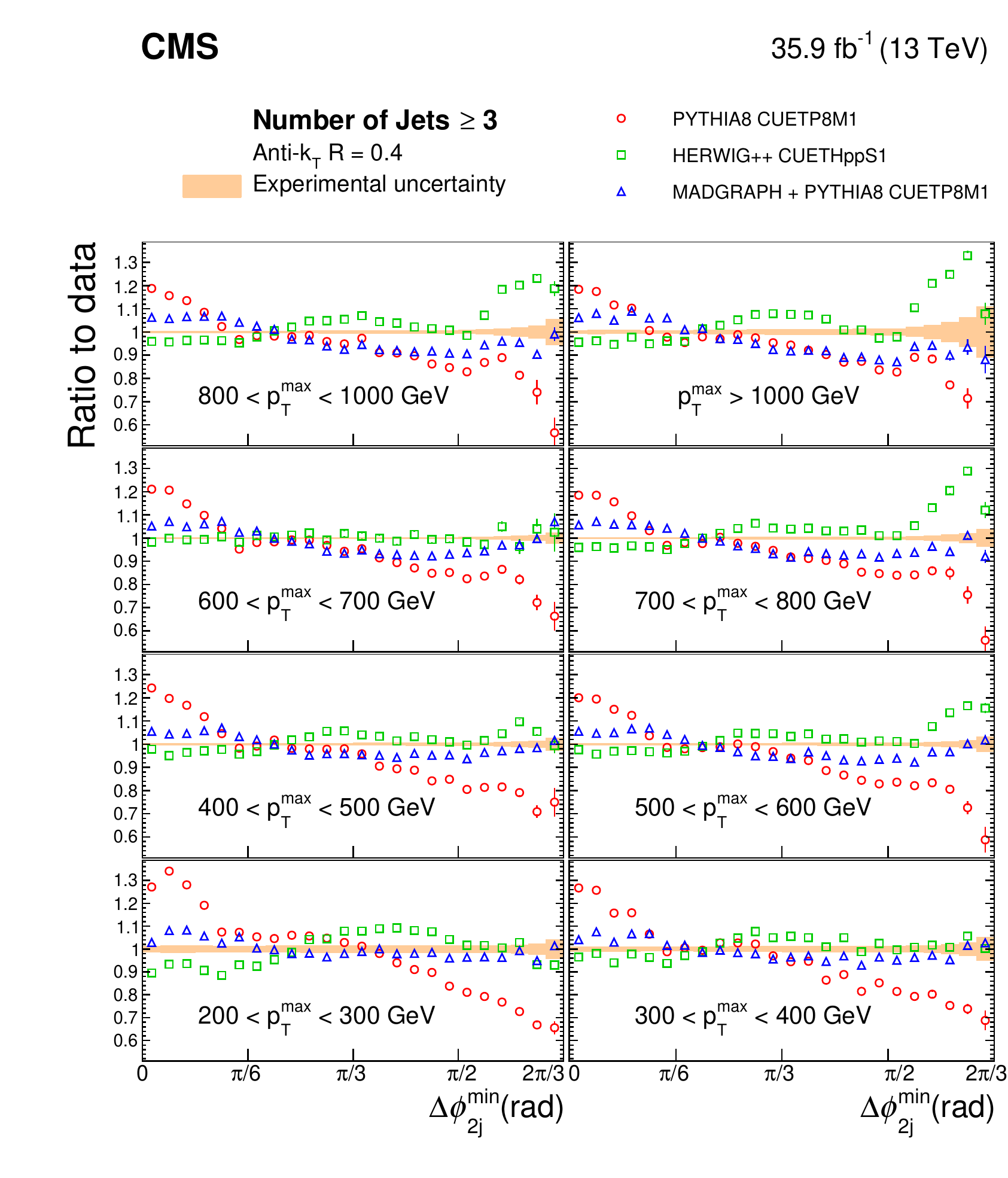}
\caption{Ratios of \PYTHIAE, \HERWIGpp, and \MADGRAPH + \PYTHIAE predictions to the normalized
inclusive 3-jet cross section differential in \dphiMinTwoJet, for all \ptmax regions.
The solid band indicates the total experimental uncertainty and
the vertical bars on the points represent the statistical
uncertainties in the simulated data.}
\label{fig:3J_ratios_MC_data_a_min2j}
\end{figure}

\begin{figure}[hbtp]
\centering
\includegraphics[width=\cmsFigWidth]{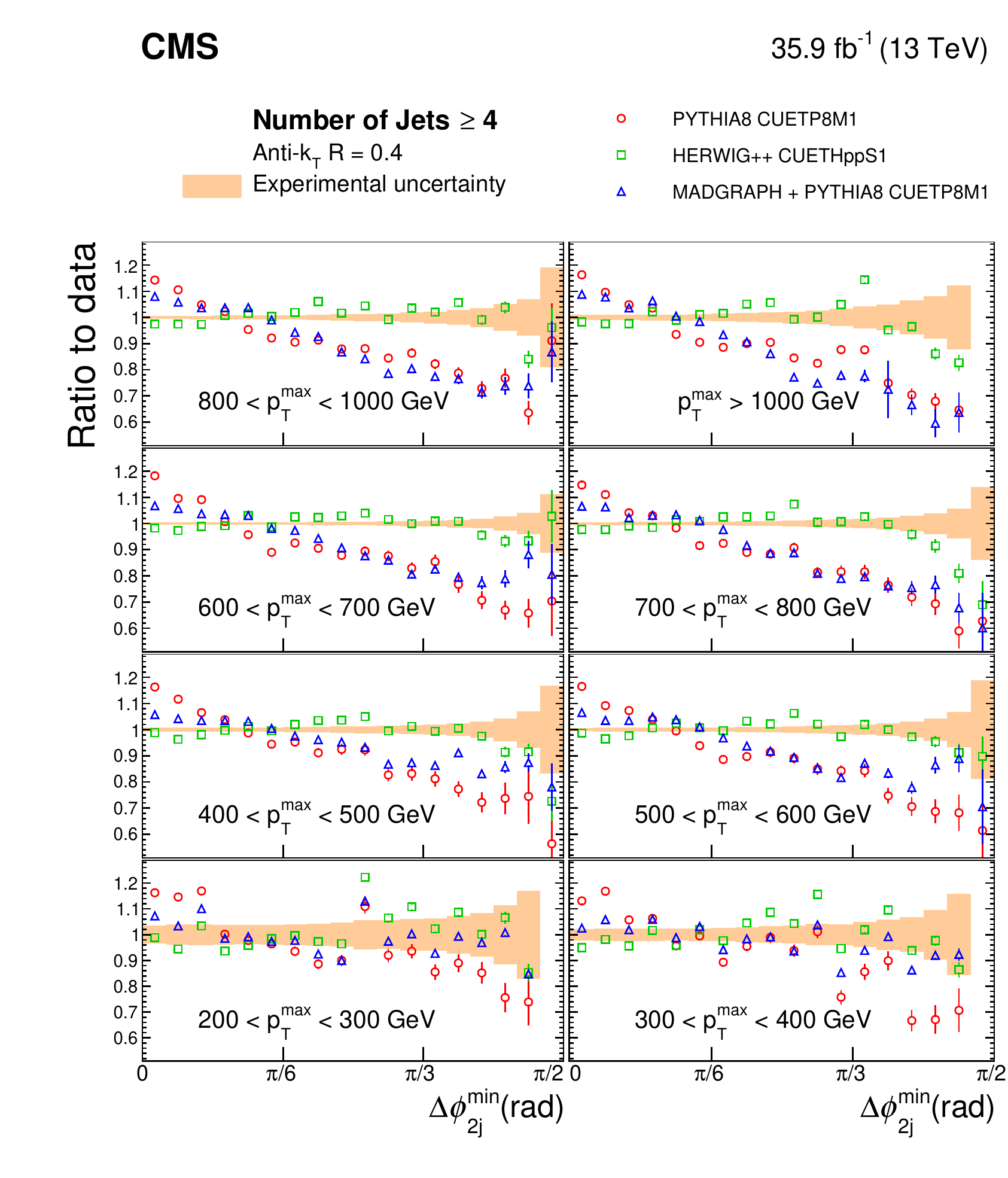}
\caption{Ratios of \PYTHIAE, \HERWIGpp, and \MADGRAPH + \PYTHIAE predictions to the normalized
inclusive 4-jet cross section differential in \dphiMinTwoJet, for all \ptmax regions.
The solid band indicates the total experimental uncertainty and
the vertical bars on the points represent the statistical
uncertainties in the simulated data.}
\label{fig:4J_ratios_MC_data_a_min2j}
\end{figure}

Figures~\ref{fig:3J_ratios_MC_data_a_min2j} and~\ref{fig:4J_ratios_MC_data_a_min2j} show, respectively, the ratios of the
\PYTHIAE, \HERWIGpp, and \MADGRAPH + \PYTHIAE event generators predictions to the normalized inclusive 3-
and 4-jet cross sections differential in \dphiMinTwoJet, for all \ptmax regions.
The \PYTHIAE event generator shows larger deviations from the measured \dphiMinTwoJet distributions in comparison
to \HERWIGpp, which provides a reasonable description of the measurement.
The \MADGRAPH generator matched to \PYTHIAE  provides a reasonable description of the measurements
in the 3-jet case, but shows deviations in the 4-jet case.

The predictions from \MADGRAPH + \PYTHIAE and \PYTHIAE are very similar for the normalized
cross sections as a function of \dphiMinTwoJet in the four-jet case.
It has been checked that predictions obtained with the \MADGRAPH matrix element
with up to 4 partons included in the calculation without contribution of the parton shower
are able to reproduce the data very well.
Parton shower effects increase the number of events with low values of \dphiMinTwoJet.

\begin{figure}[hbtp]
\centering
\includegraphics[width=\cmsFigWidth]{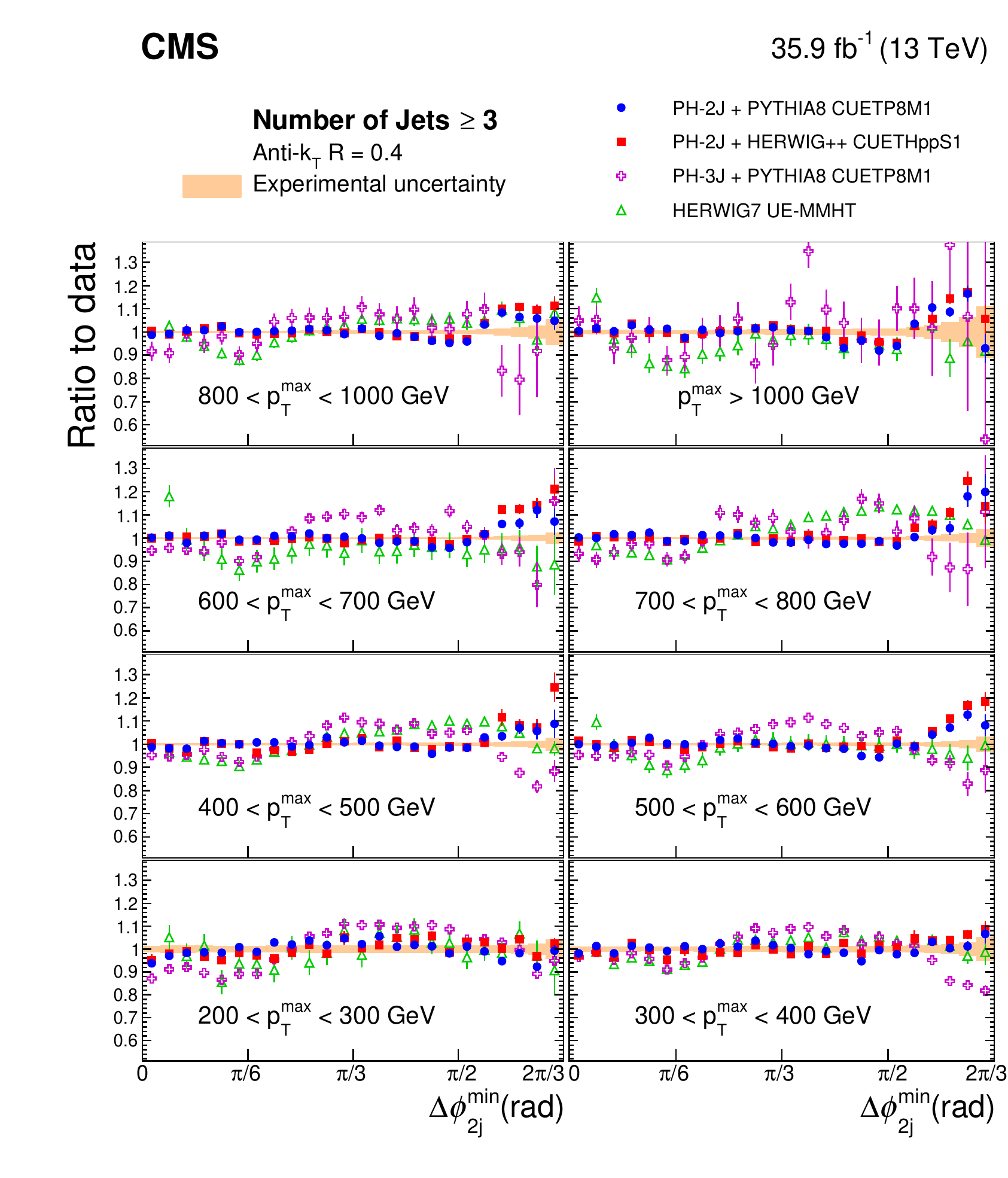}
\caption{Ratios of  \POWHEGTWOJET + \PYTHIAE, \POWHEGTWOJET + \HERWIGpp,
\POWHEGTHREEJET + \PYTHIAE, and \HERWIGSEVEN predictions to the normalized
inclusive 3-jet cross section differential in \dphiMinTwoJet, for all \ptmax regions.
The solid band indicates the total experimental uncertainty and
the vertical bars on the points represent the statistical
uncertainties of the simulated data.}
\label{fig:3J_ratios_MC_data_b_min2j}
\end{figure}

\begin{figure}[hbtp]
\centering
\includegraphics[width=\cmsFigWidth]{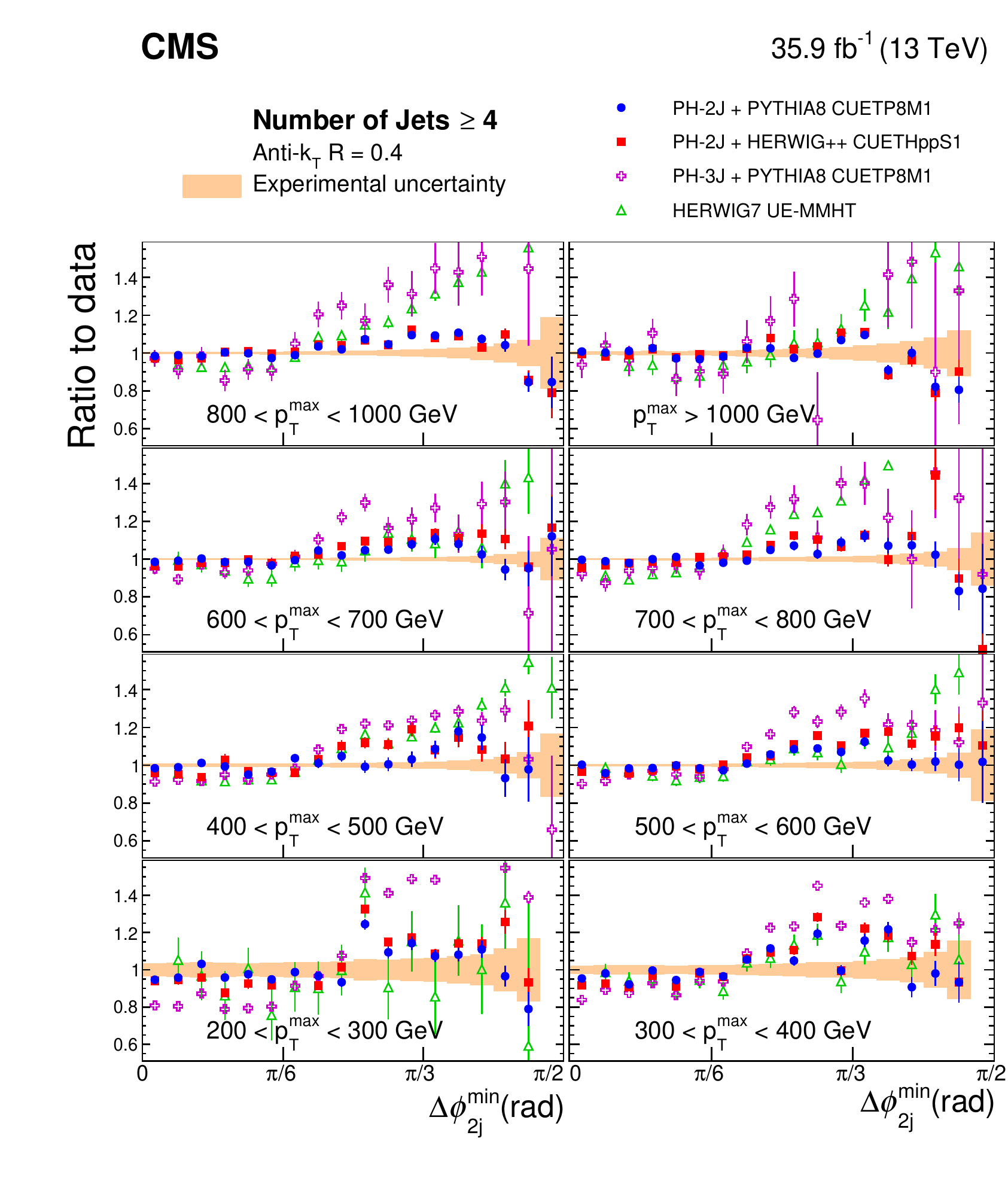}
\caption{Ratios of  \POWHEGTWOJET + \PYTHIAE, \POWHEGTWOJET + \HERWIGpp,
\POWHEGTHREEJET + \PYTHIAE, and \HERWIGSEVEN predictions to the normalized
inclusive 4-jet cross section differential in \dphiMinTwoJet, for all \ptmax regions.
The solid band indicates the total experimental uncertainty and
the vertical bars on the points represent the statistical
uncertainties of the simulated data.}
\label{fig:4J_ratios_MC_data_b_min2j}
\end{figure}

Figures~\ref{fig:3J_ratios_MC_data_b_min2j} and~\ref{fig:4J_ratios_MC_data_b_min2j} illustrate the ratios of predictions
from  \POWHEGTWOJET matched to \PYTHIAE and \HERWIGpp, \POWHEGTHREEJET + \PYTHIAE, and \HERWIGSEVEN
to the normalized inclusive 3- and 4-jet cross sections differential in \dphiMinTwoJet, for all \ptmax regions.
Due to an unphysical behavior of the \HERWIGSEVEN prediction (which has been confirmed by the \HERWIGSEVEN authors),
the first \dphiMinTwoJet and last \dphiOneTwo bins are not shown in Figs.~\ref{fig:3J_ratios_MC_data_b_12},
\ref{fig:4J_ratios_MC_data_b_12},  \ref{fig:3J_ratios_MC_data_b_min2j}, and \ref{fig:4J_ratios_MC_data_b_min2j}.
An additional uncertainty is introduced to the prediction of \HERWIGSEVEN, that is evaluated as the
difference between this prediction and the prediction when the first bin is replaced with the result from \HERWIGpp.
The additional uncertainty ranges from 2 to 10\%.
Among the three NLO dijet calculations \POWHEGTWOJET matched to \PYTHIAE or to \HERWIGpp
provides the best description of the measurements.

For the two lowest \ptmax regions in Figs~\ref{fig:4J_ratios_MC_data_a_min2j} and~\ref{fig:4J_ratios_MC_data_b_min2j},
which correspond to the 4-jet case, the measurements become statistically limited because the data used for these two regions
were collected with highly prescaled triggers with \pt thresholds of 140 and 200 \GeV (c.f. Table~\ref{tbl:TrigLumi}).

The \POWHEGTHREEJET predictions suffer from low statistical accuracy, especially in the highest interval of \ptmax, because
the same \pt threshold is applied to all 3 jets resulting in low efficiency at large \pt.
Nevertheless, the performance of the \POWHEGTHREEJET simulation on multijet observables can already be inferred by the presented predictions,
especially in the low \pt region.

The effect of parton shower uncertainties in the event generator predictions of $\dphiMinTwoJet$ is estimated to be less than 10\% over the entire phase space.

\section{Summary}
\label{sec:summary}

Measurements of the normalized inclusive 2-, 3-, and 4-jet cross sections differential in the
azimuthal angular separation \dphiOneTwo
and of the normalized inclusive 3- and 4-jet cross sections differential in the
minimum azimuthal angular separation between any two jets \dphiMinTwoJet are
presented for several regions of the leading-jet transverse momentum \ptmax.
The measurements are performed using data collected during 2016 with the CMS detector at the CERN LHC
corresponding to an integrated luminosity of 35.9\fbinv of proton-proton collisions at $\sqrt{s}=13\TeV$.

The measured distributions in \dphiOneTwo and \dphiMinTwoJet are compared with predictions from \PYTHIAE, \HERWIGpp,
\MADGRAPH + \PYTHIAE, \POWHEGTWOJET matched to \PYTHIAE and \HERWIGpp, \POWHEGTHREEJET + \PYTHIAE, and \HERWIGSEVEN event generators.

The leading order (LO) \PYTHIAE dijet event generator exhibits small deviations from the \dphiOneTwo measurements
but shows significant deviations at low-\pt in the \dphiMinTwoJet distributions.
The \HERWIGpp event generator exhibits the largest deviations of any of the generators for the \dphiOneTwo measurements,
but provides a reasonable description of the \dphiMinTwoJet distributions.
The tree-level multijet event generator \MADGRAPH in combination with \PYTHIAE for showering,
hadronization, and multiparton interactions provides a good overall description of the measurements,
except for the \dphiMinTwoJet distributions in the 4-jet case, where the generator deviates from the measurement mainly at high \ptmax.

The dijet next-to-leading order (NLO) \POWHEGTWOJET event generator deviates from the \dphiOneTwo measurements,
but provides a good description of the \dphiMinTwoJet observable.
The predictions from the three-jet NLO \POWHEGTHREEJET event generator exhibit large deviations from the measurements
and describe the considered multijet observables in a less accurate way than the predictions from \POWHEGTWOJET.
Parton shower contributions are responsible for the different behaviour of the \POWHEGTWOJET and \POWHEGTHREEJET predictions.
Finally, predictions from the dijet NLO \HERWIGSEVEN event generator matched to parton shower contributions with the MC@NLO method provide a very good description of the \dphiOneTwo
measurements, showing improvement in comparison to \HERWIGpp.

All these observations emphasize the need to improve predictions for multijet production.
Similar observations, for the inclusive 2-jet cross sections differential in \dphiOneTwo,
were reported previously by CMS \cite{bib:CMS_2} at a different centre-of-mass energy of 8\TeV.
The extension of \dphiOneTwo correlations, and the measurement of the \dphiMinTwoJet distributions
in inclusive 3- and 4-jet topologies are novel measurements of the present analysis.

\begin{acknowledgments}

We thank Simon Pl\"{a}tzer and Simone Alioli for discussion and great help on setting up, respectively, the \HERWIGSEVEN and the \POWHEGTHREEJET simulation.

We congratulate our colleagues in the CERN accelerator departments for the excellent performance of the LHC and thank the technical and administrative staffs at CERN and at other CMS institutes for their contributions to the success of the CMS effort. In addition, we gratefully acknowledge the computing centres and personnel of the Worldwide LHC Computing Grid for delivering so effectively the computing infrastructure essential to our analyses. Finally, we acknowledge the enduring support for the construction and operation of the LHC and the CMS detector provided by the following funding agencies: BMWFW and FWF (Austria); FNRS and FWO (Belgium); CNPq, CAPES, FAPERJ, and FAPESP (Brazil); MES (Bulgaria); CERN; CAS, MoST, and NSFC (China); COLCIENCIAS (Colombia); MSES and CSF (Croatia); RPF (Cyprus); SENESCYT (Ecuador); MoER, ERC IUT, and ERDF (Estonia); Academy of Finland, MEC, and HIP (Finland); CEA and CNRS/IN2P3 (France); BMBF, DFG, and HGF (Germany); GSRT (Greece); OTKA and NIH (Hungary); DAE and DST (India); IPM (Iran); SFI (Ireland); INFN (Italy); MSIP and NRF (Republic of Korea); LAS (Lithuania); MOE and UM (Malaysia); BUAP, CINVESTAV, CONACYT, LNS, SEP, and UASLP-FAI (Mexico); MBIE (New Zealand); PAEC (Pakistan); MSHE and NSC (Poland); FCT (Portugal); JINR (Dubna); MON, RosAtom, RAS, RFBR and RAEP (Russia); MESTD (Serbia); SEIDI, CPAN, PCTI and FEDER (Spain); Swiss Funding Agencies (Switzerland); MST (Taipei); ThEPCenter, IPST, STAR, and NSTDA (Thailand); TUBITAK and TAEK (Turkey); NASU and SFFR (Ukraine); STFC (United Kingdom); DOE and NSF (USA).

\hyphenation{Rachada-pisek} Individuals have received support from the Marie-Curie programme and the European Research Council and Horizon 2020 Grant, contract No. 675440 (European Union); the Leventis Foundation; the A. P. Sloan Foundation; the Alexander von Humboldt Foundation; the Belgian Federal Science Policy Office; the Fonds pour la Formation \`a la Recherche dans l'Industrie et dans l'Agriculture (FRIA-Belgium); the Agentschap voor Innovatie door Wetenschap en Technologie (IWT-Belgium); the Ministry of Education, Youth and Sports (MEYS) of the Czech Republic; the Council of Science and Industrial Research, India; the HOMING PLUS programme of the Foundation for Polish Science, cofinanced from European Union, Regional Development Fund, the Mobility Plus programme of the Ministry of Science and Higher Education, the National Science Center (Poland), contracts Harmonia 2014/14/M/ST2/00428, Opus 2014/13/B/ST2/02543, 2014/15/B/ST2/03998, and 2015/19/B/ST2/02861, Sonata-bis 2012/07/E/ST2/01406; the National Priorities Research Program by Qatar National Research Fund; the Programa Severo Ochoa del Principado de Asturias; the Thalis and Aristeia programmes cofinanced by EU-ESF and the Greek NSRF; the Rachadapisek Sompot Fund for Postdoctoral Fellowship, Chulalongkorn University and the Chulalongkorn Academic into Its 2nd Century Project Advancement Project (Thailand); the Welch Foundation, contract C-1845; and the Weston Havens Foundation (USA).

\end{acknowledgments}

\ifthenelse{\boolean{cms@external}}{}{\clearpage}

\bibliography{auto_generated}

\cleardoublepage \appendix\section{The CMS Collaboration \label{app:collab}}\begin{sloppypar}\hyphenpenalty=5000\widowpenalty=500\clubpenalty=5000\textbf{Yerevan Physics Institute,  Yerevan,  Armenia}\\*[0pt]
A.M.~Sirunyan, A.~Tumasyan
\vskip\cmsinstskip
\textbf{Institut f\"{u}r Hochenergiephysik,  Wien,  Austria}\\*[0pt]
W.~Adam, F.~Ambrogi, E.~Asilar, T.~Bergauer, J.~Brandstetter, E.~Brondolin, M.~Dragicevic, J.~Er\"{o}, M.~Flechl, M.~Friedl, R.~Fr\"{u}hwirth\cmsAuthorMark{1}, V.M.~Ghete, J.~Grossmann, J.~Hrubec, M.~Jeitler\cmsAuthorMark{1}, A.~K\"{o}nig, N.~Krammer, I.~Kr\"{a}tschmer, D.~Liko, T.~Madlener, I.~Mikulec, E.~Pree, N.~Rad, H.~Rohringer, J.~Schieck\cmsAuthorMark{1}, R.~Sch\"{o}fbeck, M.~Spanring, D.~Spitzbart, W.~Waltenberger, J.~Wittmann, C.-E.~Wulz\cmsAuthorMark{1}, M.~Zarucki
% ...
\vskip\cmsinstskip
% ...
1:~~Also at Vienna University of Technology, Vienna, Austria\\
% ...

\end{sloppypar}
\end{document}